\documentclass{article}
\usepackage{graphicx} 
\usepackage{subcaption} 

\title{Electric potential of insulated conducting objects in presence of electric charges - some exact and approximate results}
\author{Karlo Filipan$^{1,}$\thanks{karlo.flipan@unicath.hr} and Hrvoje \v{S}tefan\v{c}i\'{c}$^{1,2,}$\thanks{hrvoje.stefancic@unicath.hr} }

\date{
\centering
$^{1}$ Catholic University of Croatia, Faculty of Engineering, Ilica 244, 10000 Zagreb, Croatia \\
$^{2}$ Catholic University of Croatia, School of Medicine, Ilica 244, 10000 Zagreb, Croatia
\vspace{0.2cm}
}

\begin{document}

\maketitle

\abstract{Determination of the electric potential of insulated conducting objects is an important problem both theoretically and practically. For an insulated conducting object in the presence of external charges or charges distributed on the object surface, the problem of potential determination is reformulated using a newly introduced $J$ formalism. Using the $J$ formalism, it is shown how the electric potential can be calculated exactly for spherical objects and efficiently approximated for other object geometries using geometrical properties of the insulated conducting object. This approach does not require calculation of the surface charge distribution at the object surface of the calculation of the electric potential in the surrounding space. Properties and the performance of the approach are investigated numerically using the Robin Hood method. Possible applications of the approach based on the $J$ formalism are outlined for calculation of capacitance of conducting objects.     }

\section{Introduction}


Electrostatics is a specific limit of a dynamic electromagnetic situation in which all electric charges are static and electric fields do not vary in time \cite{Jackson,Landau,Griffiths,Reitz,Smythe}. There are numerous practically important phenomena and applications where the electric fields are not dependent on time and the electrostatic description is suitable. Some examples comprise high voltage insulation \cite{HVI}, electric propulsion \cite{propulsion}, design of microscale and nanoscale devices \cite{micronano}, electrostatic effects in materials \cite{materials} and chemistry \cite{chem1, chem2}, and various applications in agriculture \cite{agri} and industrial production in general \cite{tribo}.

Conducting objects (such as metallic electrodes) are of particular importance in electrostatic systems. When they are exposed to external electric static fields, the free electric charge in conductors redistributes at the surface of the conducting object so that its surface is equipotential, and the resulting electric field in the conductor vanishes. Finding the electric charge surface density is important both theoretically (as an element of understanding the electric field in the entire system) and practically (e.g. since the electric charge surface density is proportional to the size of the electric field close to the conducting body surface).

A particular class of conducting objects are insulated conducting objects. They are not in contact with other conducting objects, and there is no boundary condition for its surface, except the general condition that it must be equipotential. Some examples falling in this class objects comprise various floating conducting objects such as aircrafts and other aerial devices, such as drones. Friction with the air during their motion causes their charging, leading to a non-vanishing electric potential. Improving the measurement of this electric potential is a matter of ongoing research \cite{float}. As the value of electric potential is in general not known (e.g. from some boundary condition), it is problematic to use numerical methods that require the definition of boundary conditions on surfaces in the system. A method which is very well adapted to this kind of requirement, used for numerical calculations in this paper, is the Robin Hood method \cite{RH1} which is described in section \ref{sec:RHmethod}. 

In this paper we develop a novel approach towards the determination of electric potential of insulated conducting objects. In particular, we show how the geometrical properties of a conducting object can be used to determine the potential in the presence of external electric charges or electric charges distributed on the surface of the insulated conductor. A crucial advantage of this approach is that it is not necessary to calculate the electric charge distribution on the surface of the insulated conducting object or to calculate the electric potential in the space around the conductor.     

The contributions of this paper are mainly to the problem of calculating the electric potential for insulated conducting objects. Some of these contributions are theoretically important, whereas others bring new practical advantages. Specifically, the results presented in the paper comprise:  
\begin{itemize}
    \item A novel approach to the calculation of the electric potential of insulated conducting objects. This new formalism, called the $J$ formalism, is a main theoretical contribution of the paper;
    \item A new derivation of some exact analytical results for a spherical conductor obtained using the new formalism. These results present a useful validation of the new formalism;  
    \item An efficient and very good approximation for the electric potential of non-spherical insulated objects. This approximation is the main practical contribution of the paper, since the electric potential is calculated directly from the geometry or from the average potential of external point charges;   
    \item An analysis of reasons for the efficiency of the aforementioned approximation;
    \item A discussion of possible extensions of the developed formalism, such as calculation of the electric capacity of conductors of arbitrary geometry.   
\end{itemize}

The organization of the paper is the following. After the first section which gives an introduction into the problem studied in this paper, the second section brings a brief presentation of the Robin Hood method which is used for numerical calculations of electric potentials and surface charge distributions. In the third section, a novel formalism for the calculation of the electric potential of insulated conducting objects is formulated analytically. Using the developed formalism, in the fourth section exact analytical results for a spherical insulated conductor are presented. These results are complemented by numerical simulations using the RH method, which corroborate the obtained analytical results. The fifth section brings an analysis of a number of non-spherical geometries of insulated objects and external charge configurations. The main practical contribution of this paper, i.e. an efficient approximation of the equilibrium electric potential of the insulated conductor, is demonstrated in this section. The sixth section contains a detailed discussion of the formalism and approximations presented in the preceding sections, as well as possible new applications of this approach to the efficient approximation of the electric capacity of objects of arbitrary geometries. The seventh section closes the paper with a summary and conclusions.

\section{The Robin Hood method}

\label{sec:RHmethod}

One of key contributions in this paper is the demonstration that generally the average value of the electric potential at the surface of the insulated conducting object produced by the external sources is a very good approximation of the equilibrium value of the electric potential. On the other hand, the verification of this claim requires a numerical method capable of treatment of conducting objects whose electric potential is not known upfront, i.e. is not defined by some boundary condition.

The Robin Hood method \cite{RH1,RH2} is a numerical method for electrostatic problems, which is especially suitable for treating systems containing insulated conductors. In the remainder of this section, we present the idea behind the Robin Hood (RH) method and the role that non-local charge transfer plays in it. In the presentation, we focus on the RH method applied to insulted conducting objects in vacuum, although the RH method has been successfully applied to systems containing conductors at a fixed electric potential, dielectrics, etc. \cite{RH1,RH2,RH3,RH4}.

Consider an insulated conducting object. The RH method starts with a division of the body surface into small surface elements (usually triangles). An initial charge distribution is chosen, in accordance with the problem requirements (e.g. if the object is neutral, a vanishing surface charge density may be chosen). The surface charge distribution is assumed to be constant over each triangle. The electric potential is calculated for each surface element. The electric potential for a triangle with a barycenter at $\vec{x}_i$ is calculated as 
\begin{equation}
\phi(\vec{x}_i) \equiv \phi_i = \sum_{j=1}^N A_{ij} q_j \, ,
\label{eq:pot_triangle}    
\end{equation}
where 
\begin{equation}
A_{ij} = \frac{1}{\Delta S_j} \int_{S_j} k \frac{d S'}{| \vec{x}_i - \vec{x}'|}\, .
\label{eq:pot_triangle_Aij}    
\end{equation}
Here $k=\frac{1}{4 \pi \varepsilon_0}$, $\varepsilon_0$ is the permitivity of the vacuum, $S_j$ refers to the surface of triangle $j$, $\Delta S_j$ is its area and $q_j$ is its electric charge. 

The surface elements with the highest and lowest electric potential values are identified. An amount of charge is transferred from the surface element with the highest electric potential to the surface element with the lowest electric potential, so that after the charge transfer the potentials at these two surface elements are equal. The electric potential values at all other surface elements are updated for contributions coming from the change in charge at two surface elements. The cycle of identification of surface elements with the minimal and maximal values of electric potential, nonlocal charge transfer, and electric potential update is repeated until a predefined precision for the condition of equipotentiality is achieved. In the RH method, the charge of the body is explicitly conserved which makes it especially suitable for the treatment of insulated conducting objects.

For conducting objects at fixed potential or homogeneous dielectrics, precise rules of nonlocal charge transfer differ, but the main idea of the algorithm is the same. The method is easily extended to multiple objects, bearing in mind that the charge transfers are performed only between triangles belonging to the same object. The method exhibits good convergence, low computational complexity and is able to handle a very large number of surface elements \cite{RH1,RH2,RH3,RH4}.

\section{Average electric potential of an insulated conducting object in presence of charges}


For insulated conducting objects of an arbitrary shape in the electric field of external sources, the problem of determining their electric potential may be conceptually simple but practically demanding. As a response to external electric fields, the free charges in the insulated conducting object are redistributed on the surface of the object so that it becomes equipotential. The first step is finding this equilibrium charge distribution, and the second step is calculating the electric potential by the integration of contributions from all surface elements. 

In this paper, we demonstrate a much simpler way of finding the electric potential of an insulated conducting object than the one described above. That is, this potential can be calculated exactly or very well approximated by the average value of the electric potential of the external sources at the surface of the insulated conducting object. From a practical point of view, this means that to find the wanted electric potential, one does not need to calculate the surface charge density at the object, and integration of contributions from surface elements is not required as well. In the remainder of this section, we develop an analytical framework useful for the demonstration of claims given above.       

Consider an insulated conducting object in vacuum and denote its surface by $S$. In the space around the object there are various sources of external electric field. These sources are assumed not to change under the influence of the surface charge distribution that appears on the surface of the insulated conducting object. This requirement on the external field is fulfilled for a fixed distribution of point charges. Our main interest will then be aimed at configurations of fixed point charges around an insulated conducting object. For a point $P \in S$ at a position $\vec{x}$, it is possible to express the value of the electric potential as the sum of the electric potential produced by external sources and contributions of charges at the surface of the object:   
\begin{equation}
\phi (\vec{x})= \phi_{ext} (\vec{x}) + \int_{S} k \frac{\sigma (\vec{x'})}{|\vec{x}-\vec{x'}|} d S' \, . 
    \label{eq:potential}
\end{equation}
Here, $\phi (\vec{x})$ denotes the electric potential at the point $P$, $\phi_{ext} (\vec{x})$ denotes the electric potential of external sources, and $\sigma (\vec{x'})$ is a surface charge density at the surface $S$.

For any quantity $X$ defined at the surface of the insulated conductor, we define an average $\langle X \rangle$ as
\begin{equation}
\langle X \rangle \equiv \frac{1}{S} \int_{S} X d S \, .
\label{eq:averaging}    
\end{equation}
Taking $ \frac{1}{S} \int_{S}  d S$ of the expression (\ref{eq:potential}), one obtains
\begin{equation}
\langle \phi \rangle = \langle \phi_{ext} \rangle + \frac{1}{S} \int_S d S \int_S d S' k \frac{\sigma(\vec{x'})}{|\vec{x}-\vec{x'}|} \, ,
\label{eq:avgpot1}
\end{equation}
which can be rewritten as
\begin{equation}
\langle \phi \rangle = \langle \phi_{ext} \rangle + \frac{1}{S} \int_S d S' \sigma(\vec{x'}) \int_S d S \, k \frac{1}{|\vec{x}-\vec{x'}|} \, .
\label{eq:avgpot2}
\end{equation}

Let us further introduce a geometric quantity $J(\vec{x'})$ defined as
\begin{equation}
J(\vec{x'}) \equiv \int_S d S \, k \, \frac{1}{ |\vec{x}-\vec{x'}|} \, .
\label{eq:Jdef}    
\end{equation}
Although $J(\vec{x'})$ could be understood as a purely geometrical quantity, it is also possible to interpret it as a potential at $\vec{x'}$ produced by the unit surface charge distribution on the surface $S$. Using (\ref{eq:Jdef}), the expression (\ref{eq:avgpot2}) can be written as
\begin{equation}
\langle \phi \rangle = \langle \phi_{ext} \rangle + \frac{1}{S} \int_S d S' \sigma(\vec{x'}) J(\vec{x'}) \, .
\label{eq:potJ}    
\end{equation}
This expression can be further reorganized to give
\begin{equation}
\langle \phi \rangle = \langle \phi_{ext} \rangle + \langle \sigma \rangle \langle J \rangle + \frac{1}{S} \int_S d S' (\sigma(\vec{x'})-\langle \sigma \rangle) (J(\vec{x'}) - \langle J \rangle) \, ,
\label{eq:potJext}    
\end{equation}
where $\langle \sigma \rangle = \frac{Q}{S}$ and $Q$ denotes the charge of the insulated conductor.

The expressions (\ref{eq:potJ}) and (\ref{eq:potJext}) allow for a novel view of the problem of calculating the electric potential of the insulated conductor. The third term in (\ref{eq:potJext}) proves to be small in many practical cases and the potential can be calculated only using the first two terms in this expression. If the insulated conductor has a spherical geometry, exact results follow from first two terms and the third term vanishes. For objects which are non-spherical, keeping only first two terms in  (\ref{eq:potJext}) provides an efficient approximation scheme which is the better the object is closer to the spherical form. In the following sections, the aforementioned claims are elaborated and illustrated. 

It should be stressed that this approach is also useful for important cases where there are no external charges, but some charge $Q$ is distributed on an insulated conducting object of an arbitrary geometry. A practical example of such a situation is an airplane charged by friction with the air \cite{float}.

As the entire approach in this paper is based on the function $J(\vec{x})$ for the insulated conducting object, we call this approach {\bf $J$ formalism}, and refer to it as such hereafter.

\section{Spherical objects}

\label{sec:sphere}

As in many problems involving spherical objects, the high level of symmetry in spherical geometries simplifies the treatment and provides analytical tractability in our case as well. An exact analytical expression for the electric potential of the insulated conducting sphere can be obtained in the $J$ formalism, as demonstrated in the next subsection.    

\subsection{Exact analytical results}

The integral expression for $J (\vec{x'})$ also exhibits spherical rotational symmetry, which leads to the conclusion that its value is the same at any point of the sphere's surface. In particular,
\begin{equation}
J (\vec{x'})= const \equiv J \, .
\label{eq:Jsphere}    
\end{equation}
This conclusion can be confirmed by the explicit calculation of the $J$ value for a sphere of a radius $R$:
\begin{equation}
J=4 \pi  k R \, .
\label{eq:Jspherevalue}    
\end{equation}
Inserting this result in (\ref{eq:avgpot2}) it is evident that the integral in (\ref{eq:avgpot2}) vanishes and the expression for $\langle \phi \rangle$ is considerably simplified:
\begin{equation}
\langle \phi \rangle = \langle \phi_{ext} \rangle + \frac{J Q}{4 \pi R^2} \, ,
\label{eq:phisphere}    
\end{equation}
or, using the explicit value of $J$ given in (\ref{eq:Jspherevalue})
\begin{equation}
\langle \phi \rangle = \langle \phi_{ext} \rangle + k \frac{Q}{R} \, .
\label{eq:phispherenext}
\end{equation}
This expression can be applied to a situation where the external potential is produced by
a point charge $q$ at the distance $a$ from the center of the sphere. In this case, the average value of the external electric potential can be calculated analytically and it amounts to
\begin{equation}
\langle \phi_{ext} \rangle =  k \frac{q}{a} \, .
\label{eq:phiextpoint}
\end{equation}
This result can be compared with a well-known result, also called {\em the mean value theorem} \cite{Jackson}. It says that in space without electric charge, the potential at the center of a sphere equals the average of the potential over the sphere centered at that point. 

Based on the principle of superposition, the result for a single external point charge obtained above can be generalized to a configuration of $n$ charges $q_i$ at positions $\vec{a}_i$, $i=1,2,\dots,n$. In this general situation, the average electric potential produced by the external sources is
\begin{equation}
\langle \phi_{ext} (q_1, \vec{a}_1; q_2, \vec{a}_2; \dots q_n, \vec{a}_n) \rangle = k  \sum_{i=1}^n \frac{q_i}{a_i} \, ,
\end{equation}
which finally leads to
\begin{equation}
\langle \phi \rangle =  k \sum_{i=1}^n \frac{q_i}{a_i} + k \frac{Q}{R} \, .
\label{eq:phispherefinal}    
\end{equation}

It is important to stress that the results for a conducting insulated sphere in the presence of point charges are not new, that is, they could be obtained from the available analytical solutions obtained by the method of images. The novel contribution, specified in Eq. (\ref{eq:phispherenext}), says that for a spherical geometry, details of the surface charge distribution are not important in determining the electric potential of the insulated sphere. Apart from the charge of the sphere, the only relevant quantity is the average of the external electric potential over the surface of the sphere. Although Eq. (\ref{eq:phispherenext}) was applied to situations where $\phi_{ext}$ is produced by point charges, this expression is also applicable to other fixed sources of electric potential, such as static continuous distributions of charge. Practical implications of this results are considerable: if we put a conducting insulated sphere in the field of any static charge distribution, to determine its electric potential it is not necessary to calculate the charge distribution on the sphere; it is sufficient to find the average value of the external electric potential over the surface of the sphere.    


\begin{figure}[ht!]
\centering
\includegraphics[width=0.9\textwidth]{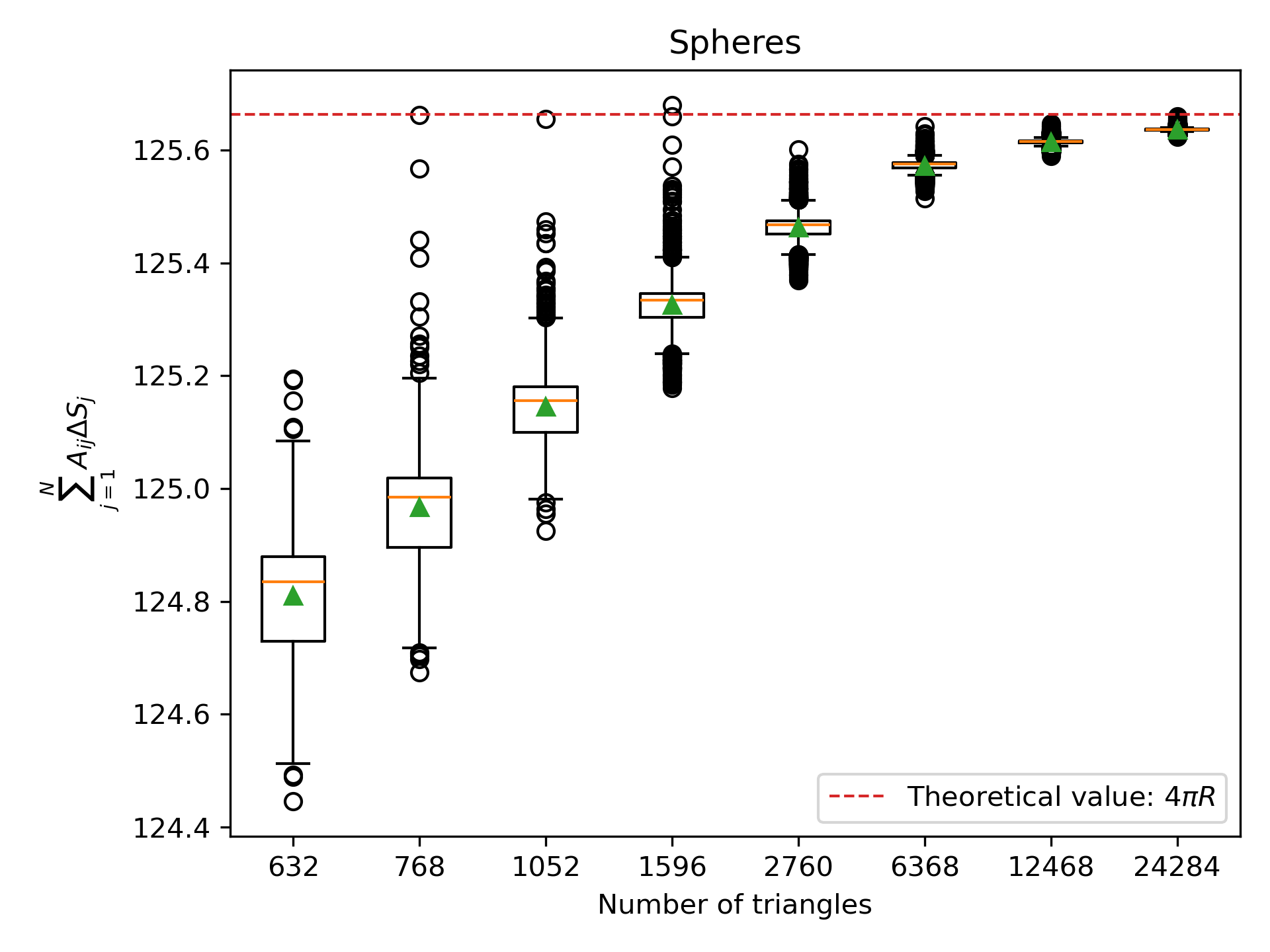}
\caption{Distributions of $J(\vec{x}_i)$ for sphere of radius $R=10$ for different mesh refinements: average approaches theoretical value of $4 \pi R$ (for $k=1$) for more refined meshes. The green triangle denotes the mean value and the orange line the median value. The orange dashed line corresponds to the theoretical value for the $J$ value of a sphere of radius $R$.}
\label{Fig:sphere_J}
\end{figure}

\subsection{Numerical results}


The findings of the previous subsection can be further corroborated by numerical calculations using the RH method. Our primary interest is to show that the $J$ formalism produces very good approximations of the electric potential values. To this end, in all numerical calculations, we adopt $k=1$ and give distances, charge densities, and potentials in arbitrary units. 

In the scope of the RH method, the following connection between the function $J$ at a barycenter of a triangle $i$ and the quantities $A_{ij}$ can be established for any surface geometry:
\begin{equation}
J(\vec{x}_i) = \sum_{j=1}^N A_{ij} \Delta S_j \, .
\label{eq:JvsA}
\end{equation}

 The values of $J(\vec{x}_i)$ values for all triangles are calculated for various dicretizations of a sphere and the dependence of $J$ value distributions on number of triangles is presented in Fig. \ref{Fig:sphere_J}. From the Figure it is clear that as the discretization is more detailed (i.e. for a bigger number of triangles), the distribution of $J$ values has a smaller dispersion and its mean and median values are closer to the theoretical value of $4 \pi R$. These findings are in line with the theoretical expectation of a constant $J$ over the surface of a sphere. 

\begin{figure}[ht!]
\centering
\begin{tabular}{cc}
\subfloat[]{
\includegraphics[width=0.5\textwidth]{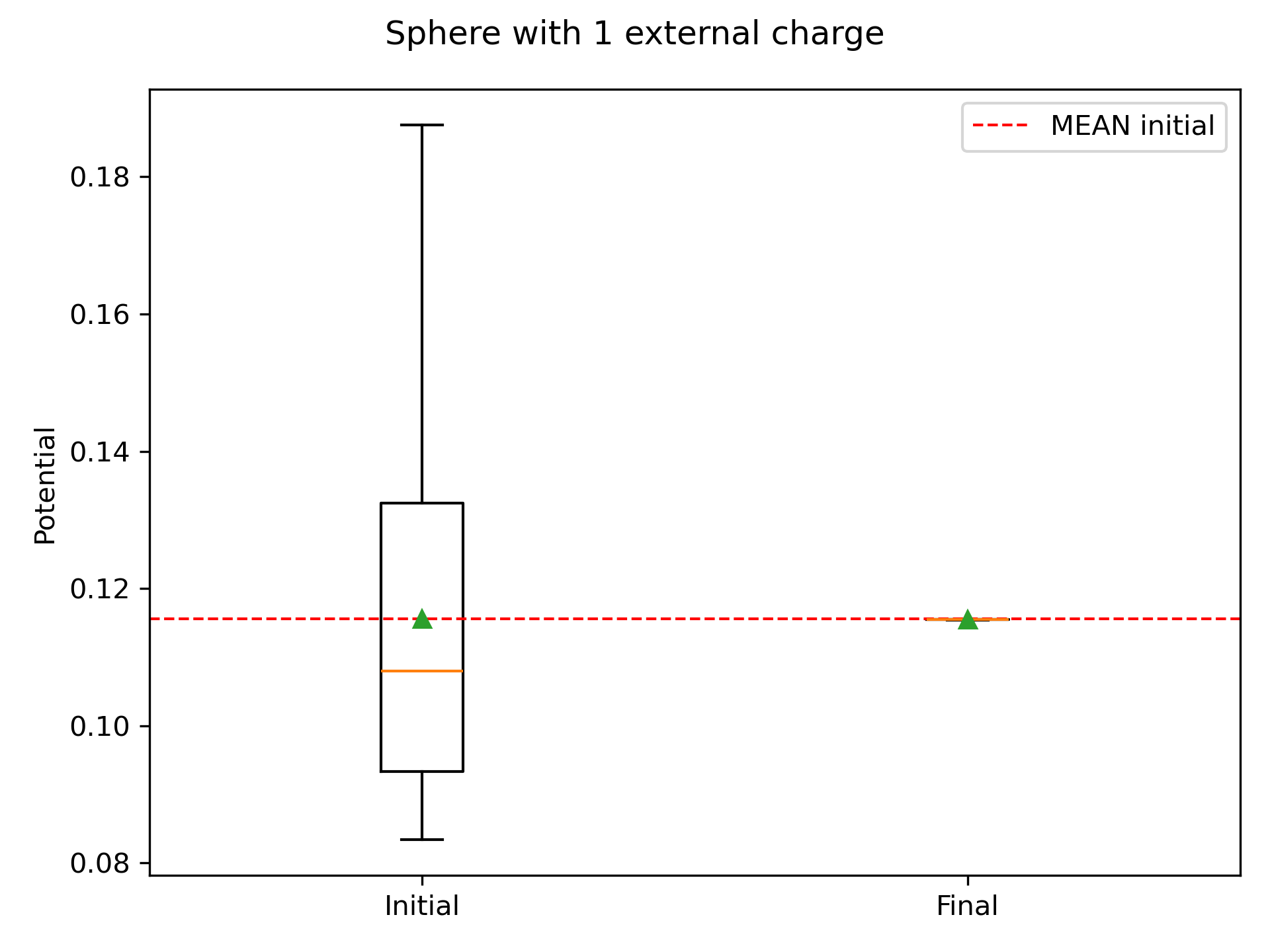}
} &
\subfloat[]{
\includegraphics[width=0.5\textwidth]{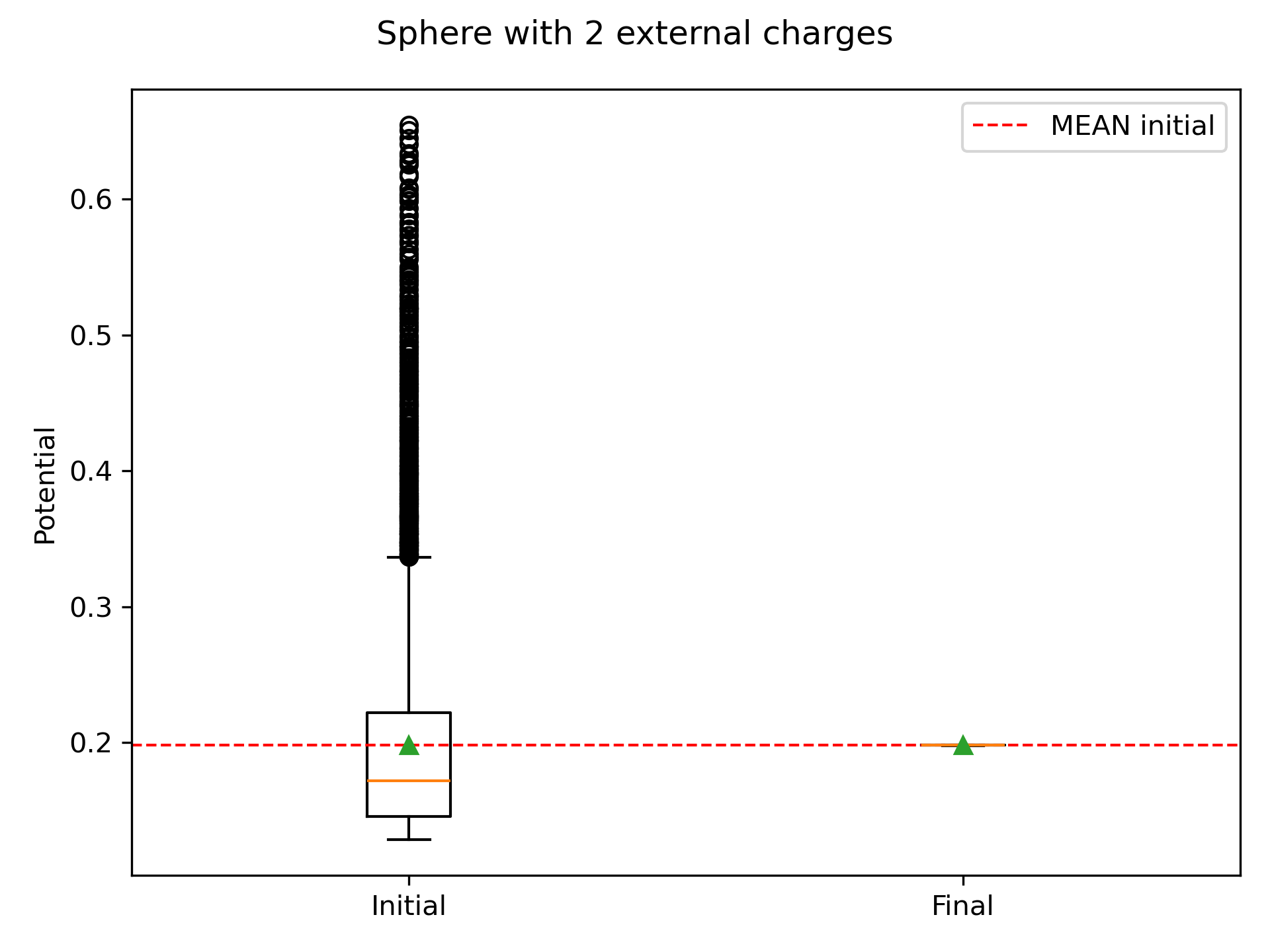}
} \\
\subfloat[]{
\includegraphics[width=0.5\textwidth]{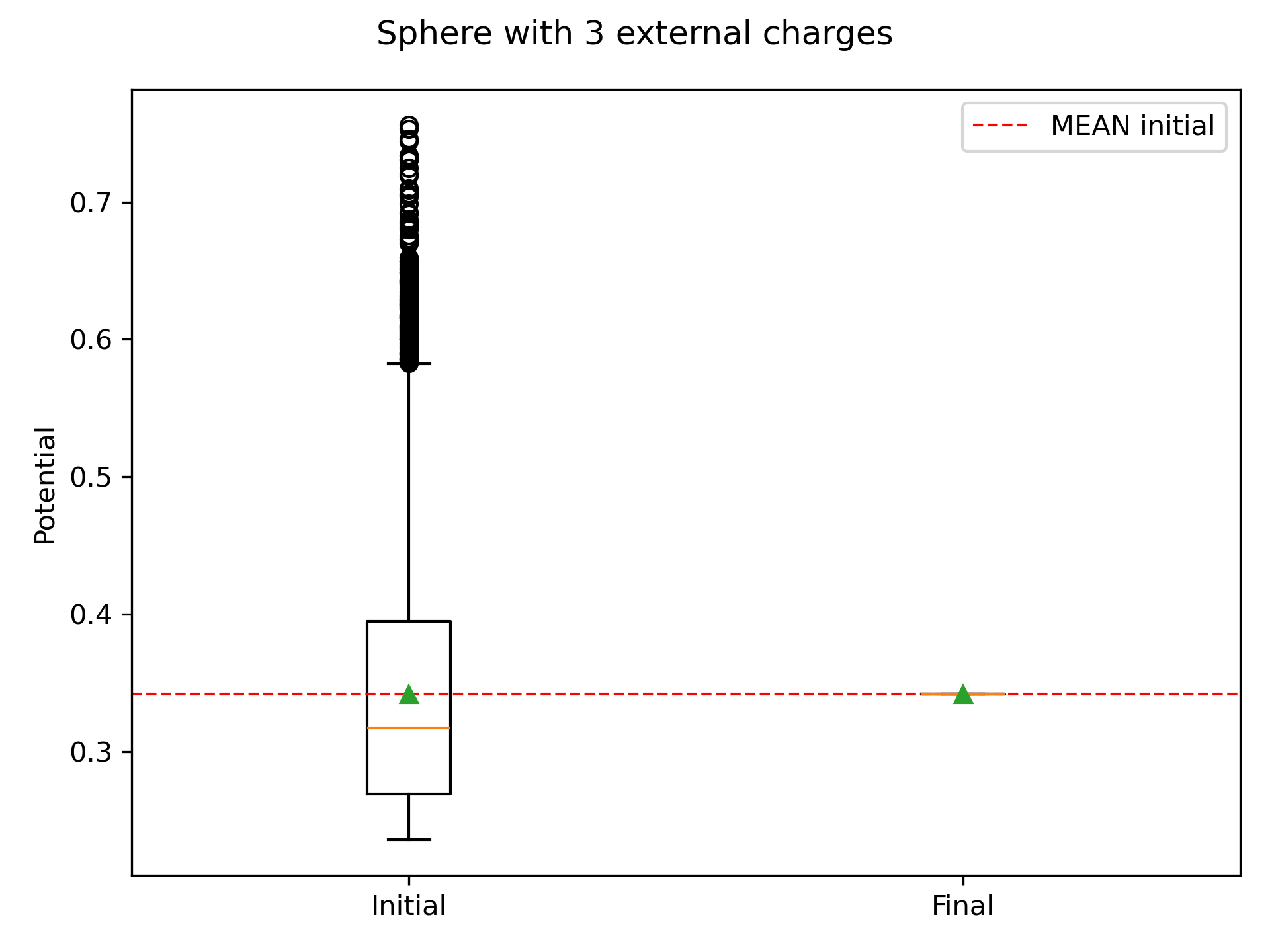}
} &
\subfloat[]{
\includegraphics[width=0.5\textwidth]{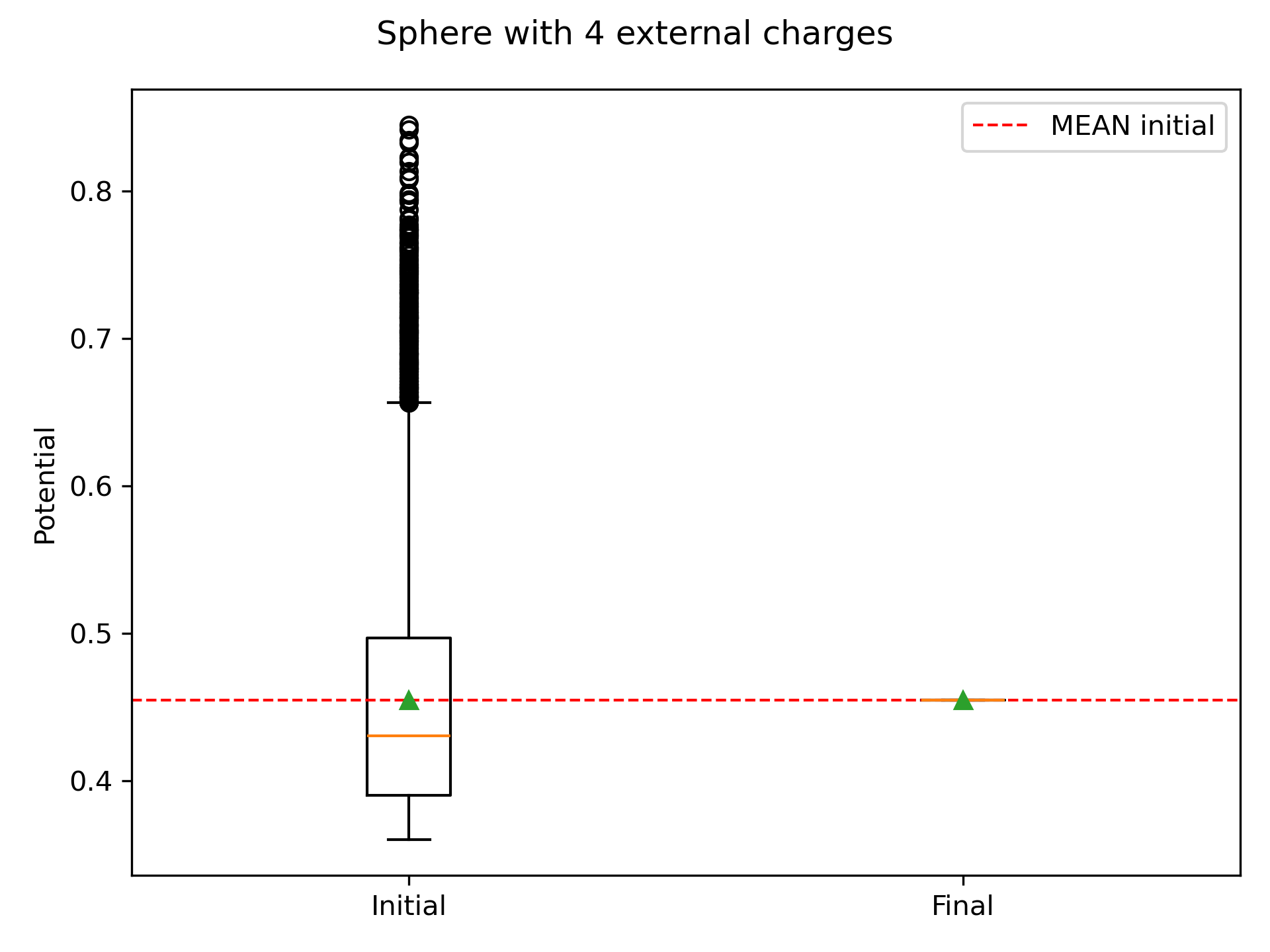}
}
\end{tabular}
\caption{Comparison of the initial and final potential distributions for the sphere of radius $R=10$, centered at $(0, 0, 0)$ consisting of $6368$ triangles. Four configurations with multiple external point charges are presented: a) a single charge of value $3$ at the position $(15, 15, 15)$; b) two charges - configuration in a) with added charge of value $1$ at the position $(7, 7, 7)$; c) three charges - configuration in b) with added charge of value $2$ at the position $(-9, 8, -7)$; d) four charges - configuration in c) with added charge of value $4$ at the position $(-15, -20, -25)$.}
\label{Fig:sphere_potential}
\end{figure}

For a spherical geometry, the average potential of the external charges should be equal to the full value of the electric potential for a neutral insulated conducting sphere. In Fig. \ref{Fig:sphere_potential} we present results for four configurations: an insulated sphere surrounded by one, two, three, and four point charges. For each of the configurations, the distribution of the electric potential at the triangles before and after the RH calculation is presented. In the RH calculation, the initial surface charge density is zero. From the plots it is evident that the initial mean value of the electric potential equals the final equilibrium value of the potential.  

\begin{figure}[ht!]
\centering
\subfloat[]{
\includegraphics[width=0.5\textwidth]{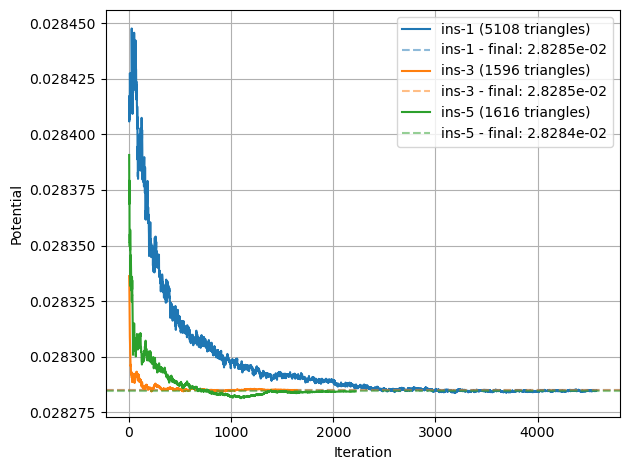}
}
\subfloat[]{
\includegraphics[width=0.5\textwidth]{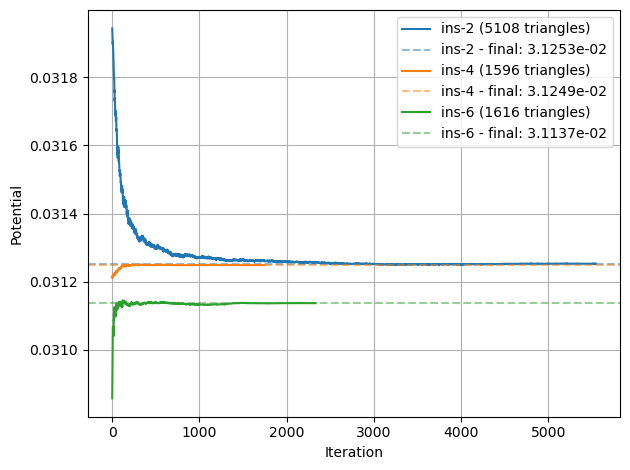}
}
\caption{Average potential across iterations for different spheres centered at $(0, 0, 0)$: two spheres are of radius 10 but with different triangulations (5108 vs 1596 triangles) and the third sphere is of radius 30 (1616 triangles). In a) external unit charge is placed at $(-15, -20, -25)$ and in b) external unit charge at $(0, 0, 32)$. 
}
\label{Fig:sphere_iterations}
\end{figure}

In Fig. \ref{Fig:sphere_iterations}, we present how the average electric potential over the spherical surface depends on the number of RH iterations. An external point charge is located at a fixed distance from the center of the sphere and spheres of different radii and triangulations are considered. As the sphere is initially not charged, the average initial potential corresponds to the average electric potential produced by the external charge. The average electric potential converges to the final value, which differs very little from the initial value (at the level of the numerical precision of the RH method), in accordance with the theoretical predictions of section \ref{sec:sphere}. From the plots in this Figure, it is clear that the final potential is not dependent on the sphere radius, as obtained in (\ref{eq:phispherefinal}).

\begin{figure}[ht!]
\centering
\subfloat[]{
\includegraphics[width=0.5\textwidth]{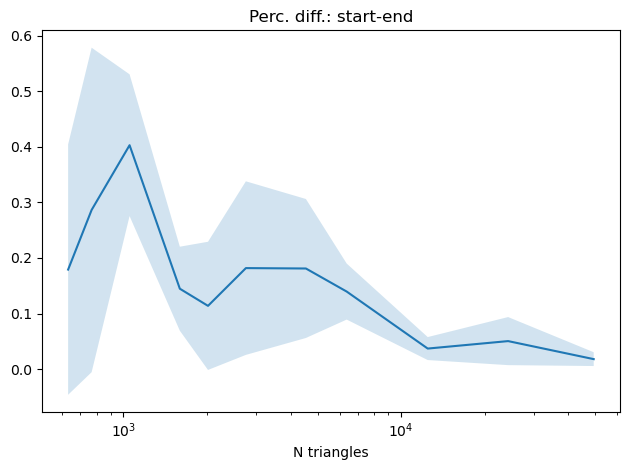}
}
\subfloat[]{
\includegraphics[width=0.5\textwidth]{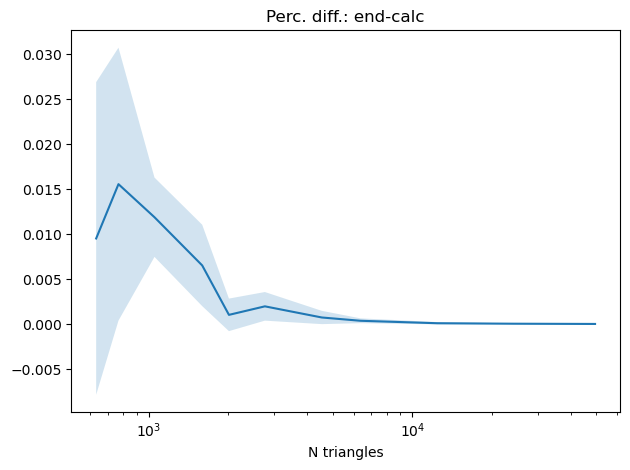}
}
\caption{Average and standard deviation of percentage differences: a) between initial and final average potential; b) between final average potential and analytical solution. The distributions are obtained for 11 spheres of the same radius with different number of triangles and for each sphere a unit charge is positioned at 7 different points in space.}
\label{Fig:sphere_triangles}
\end{figure}

For a sphere, the theoretical results show that the difference between initial and final value of the average electric potential should be zero. It is expected that the deviation of this difference from zero is a consequence of numerical approximations (such as triangulation of the spherical surface) and that with the refinement of the triangular mesh used to represent the spherical surface this deviation should be reduced. To test this expectation, we consider a number of meshes with different number of triangles and calculate the difference for several charge positions. In Fig. \ref{Fig:sphere_triangles}, we present how the distribution of the difference depends on the number of triangles used in the discretization of the surface. It is evident that it converges toward zero as the number of triangles increases. Additionally, in line with expectations, the final numerical result is closer to the analytical solution for more refined meshes.


\section{Non-spherical objects}


\begin{figure}[ht!]
\centering
\includegraphics[width=0.9\textwidth]{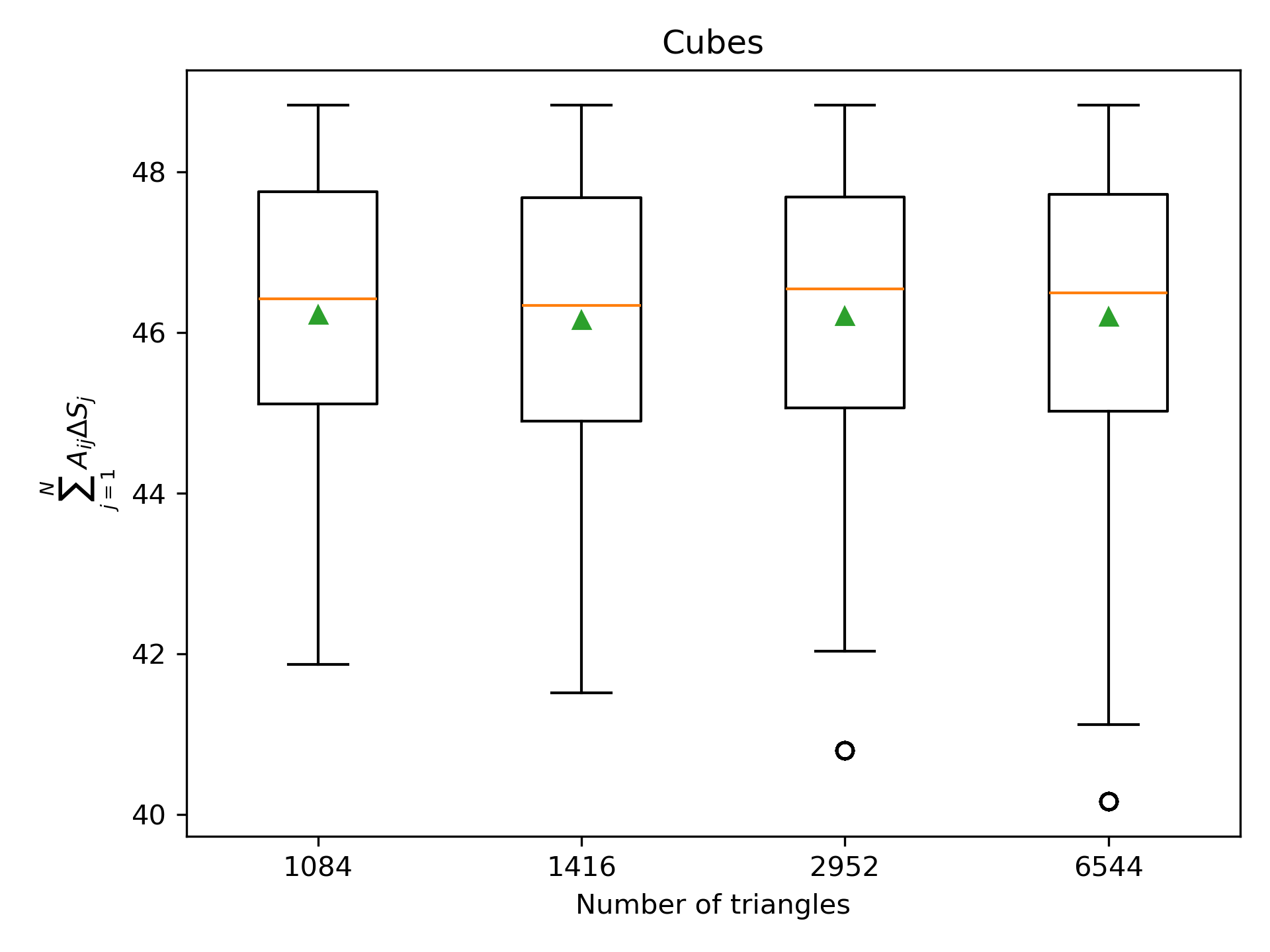}
\caption{Distributions of $J(\vec{x}_i)$ for a cube of side 5 for different mesh refinements.}
\label{Fig:cubes_J}
\end{figure}

\begin{figure}[ht!]
\centering
\includegraphics[width=0.9\textwidth]{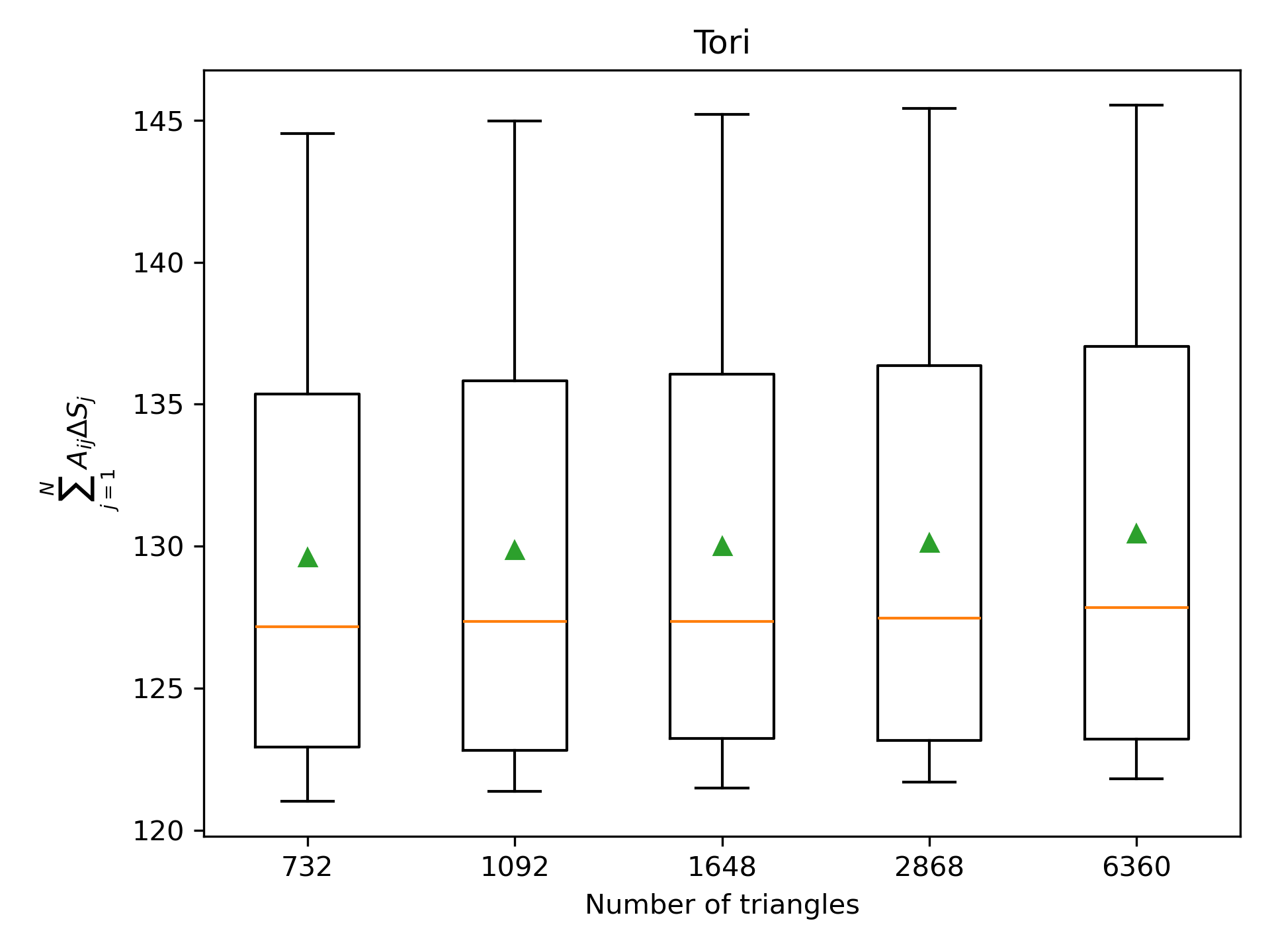}
\caption{Distributions of $J(\vec{x}_i)$ for a torus with radii $R = 10$ and $r = \frac{10}{3}$ for different mesh refinements.}
\label{Fig_tori_J}
\end{figure}

\subsection{Numerical results}

For non-spherical objects the values of $J(\vec{x}_i)$ are not all equal, and they can be found in some interval depending on precise details of geometry. Therefore, it is not expected that the refinement of surface discretization will reduce the dispersion of $J$ values for the object. Indeed, in Fig. \ref{Fig:cubes_J} for cubes and in Fig. \ref{Fig_tori_J} for tori, one can see that the distribution of $J$ values is not changed in any systematic way as the number of triangles approximating the object surface increases. 

\begin{figure}[ht!]
\centering
\begin{tabular}{cc}
\subfloat[]{
\includegraphics[width=0.5\textwidth]{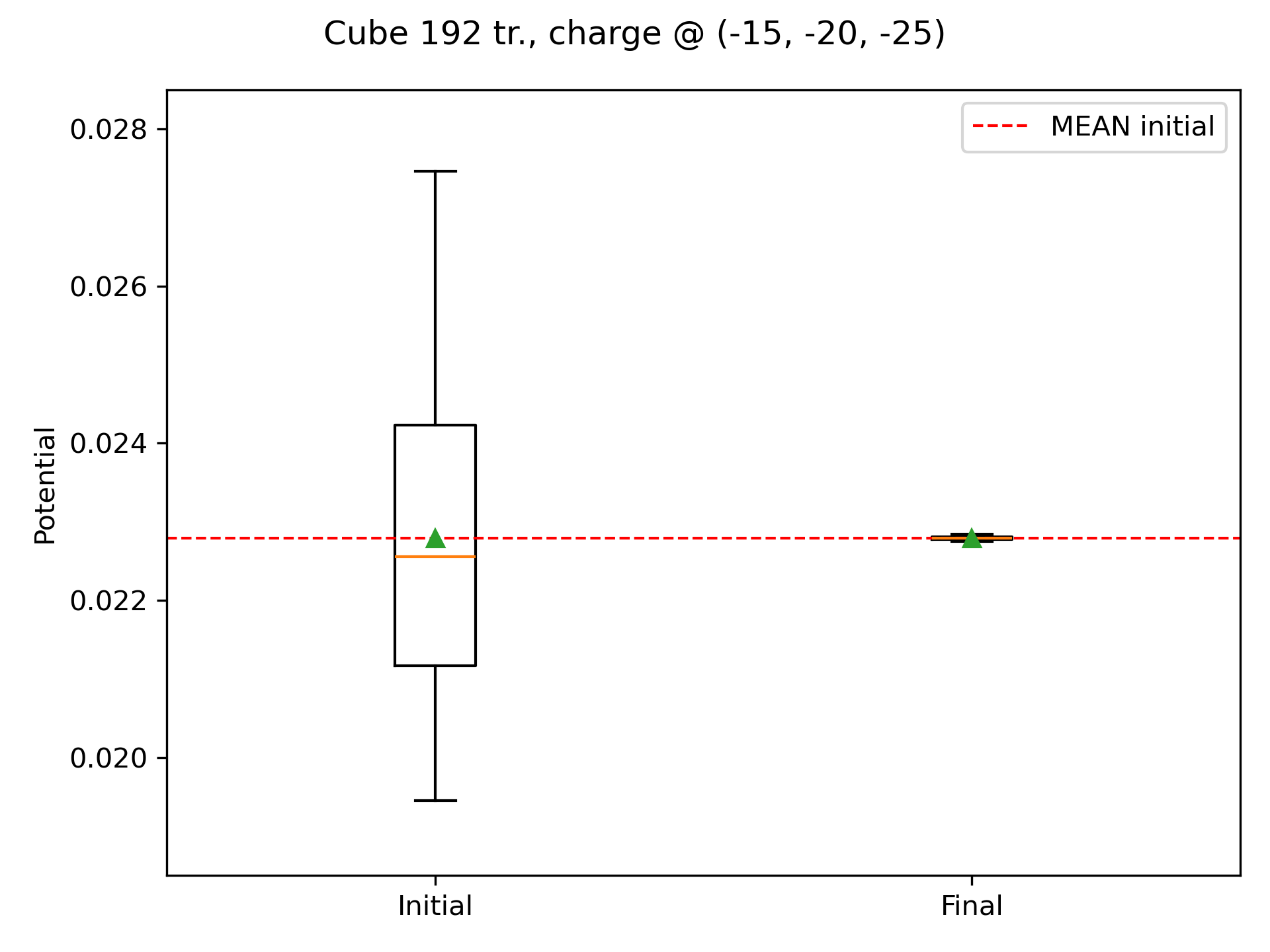}
} &
\subfloat[]{
\includegraphics[width=0.5\textwidth]{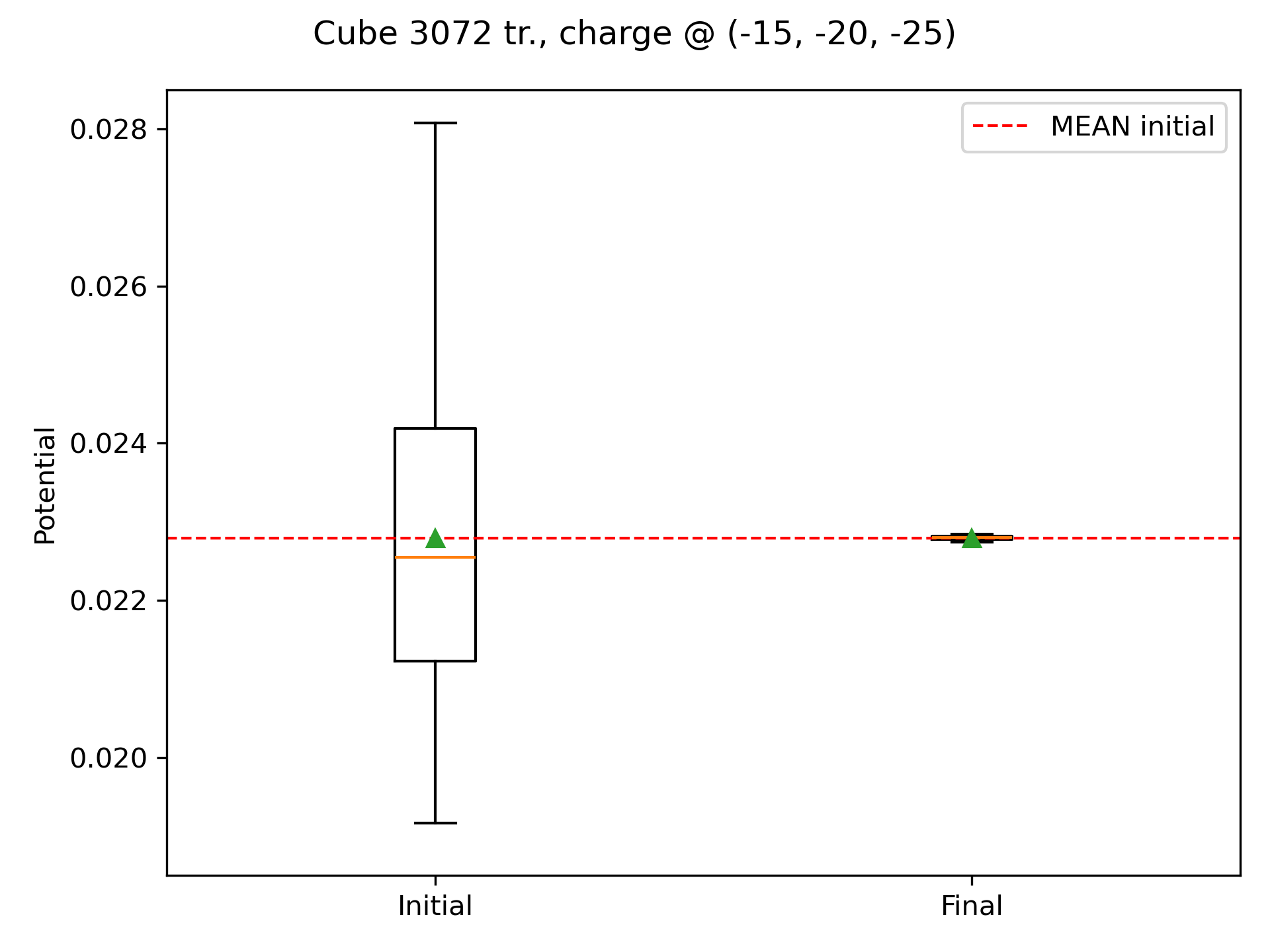}
} \\
\subfloat[]{
\includegraphics[width=0.5\textwidth]{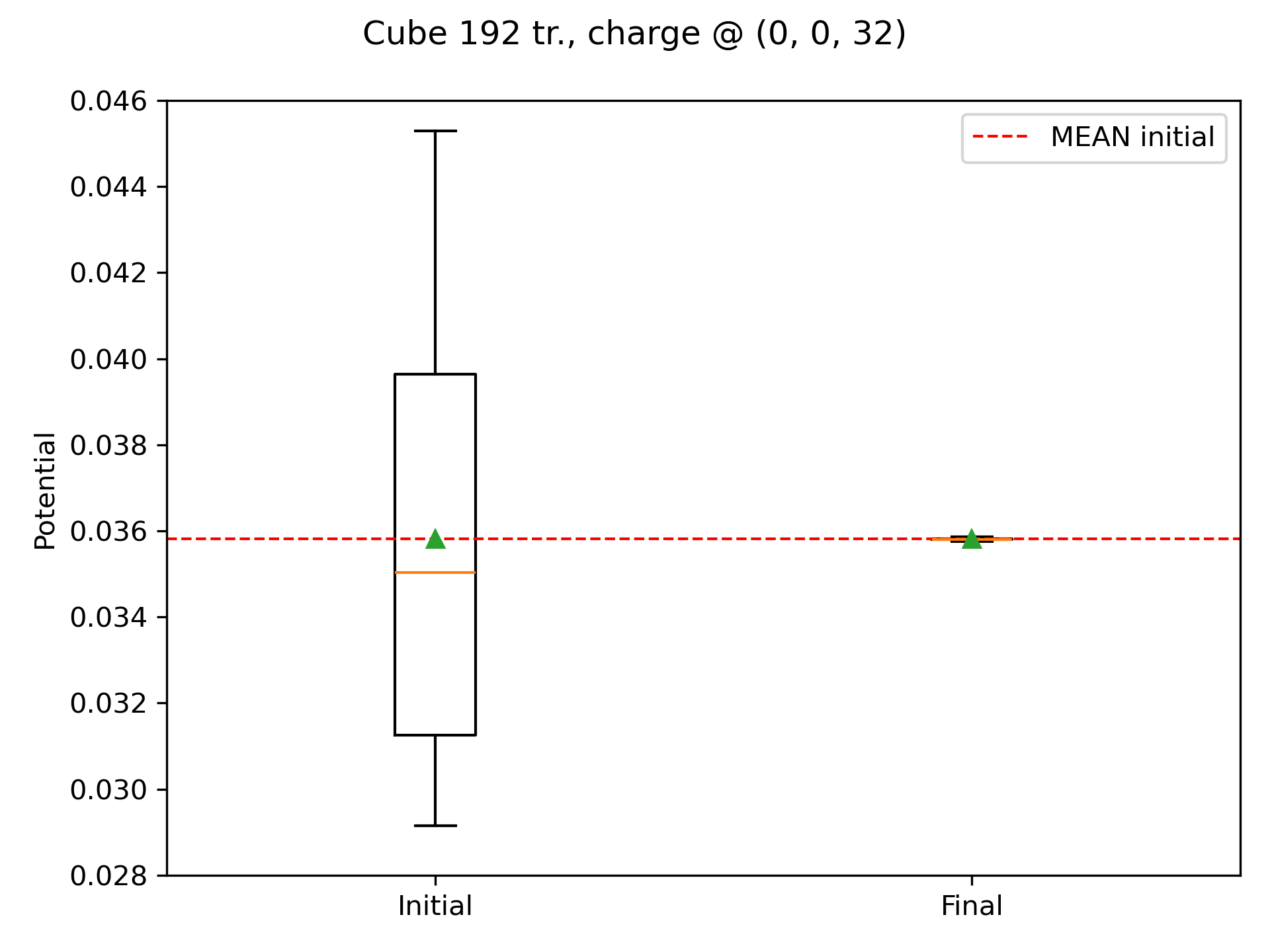}
} &
\subfloat[]{
\includegraphics[width=0.5\textwidth]{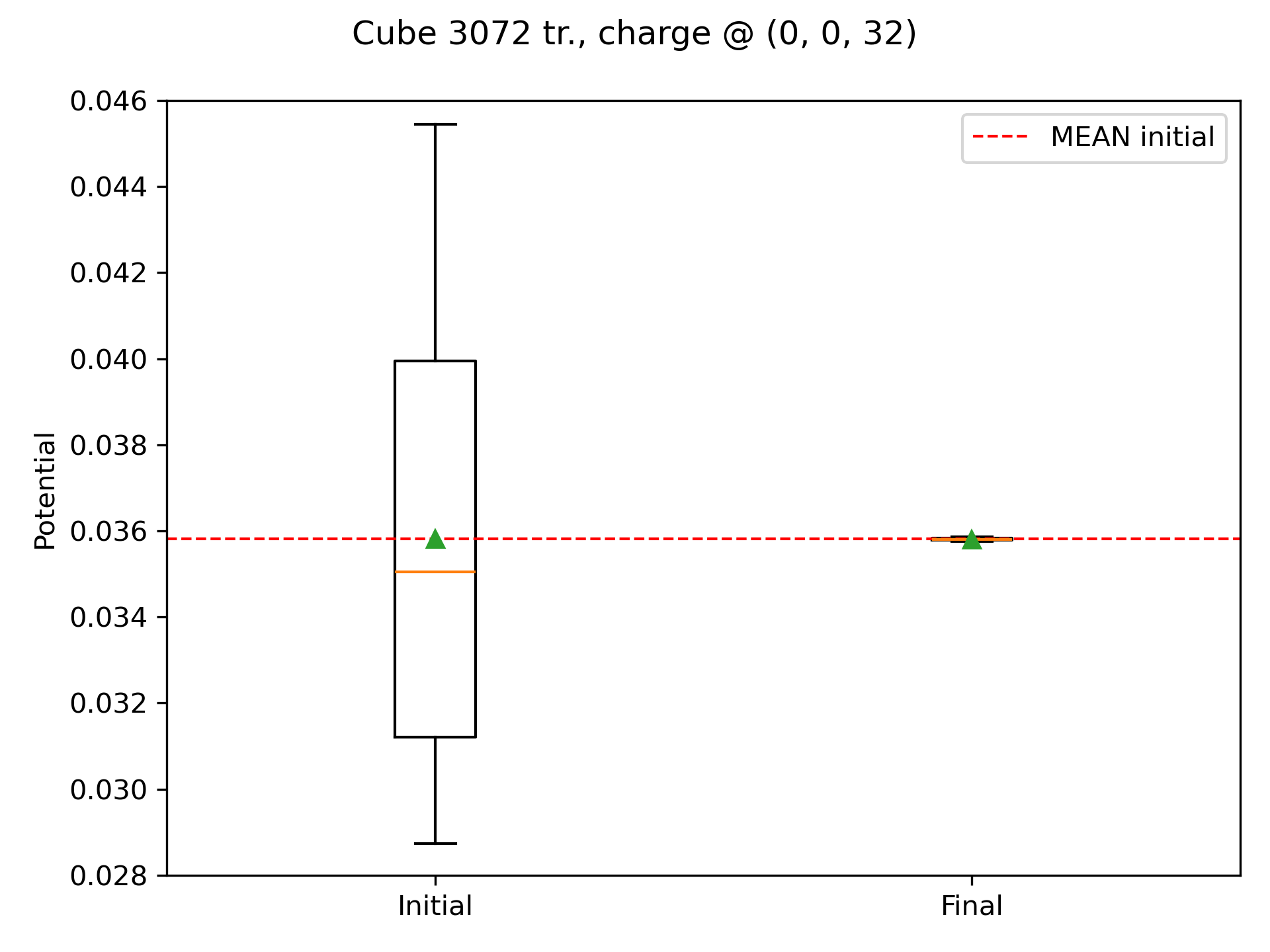}
}
\end{tabular}
\caption{Initial and final potential distributions for examples with a single unit point charge located outside the insulated neutral cube having one vertex at the position $(0, 0, 0)$ and sides of length 10 in direction of $(x, y, z)$ axes: a) the cube discretization into 192 triangles with a point charge at $(-15, -20, -25)$, b) the cube discretization into 3072 triangles with charge at $(-15, -20, -25)$, c) the cube discretization into 192 triangles with charge at $(0, 0, 32)$, d) the cube discretization into  3072 triangles with charge at $(0, 0, 32)$.}
\label{Fig:potential_cubes}
\end{figure}

\begin{figure}[ht!]
\centering
\begin{tabular}{cc}
\subfloat[]{
\includegraphics[width=0.5\textwidth]{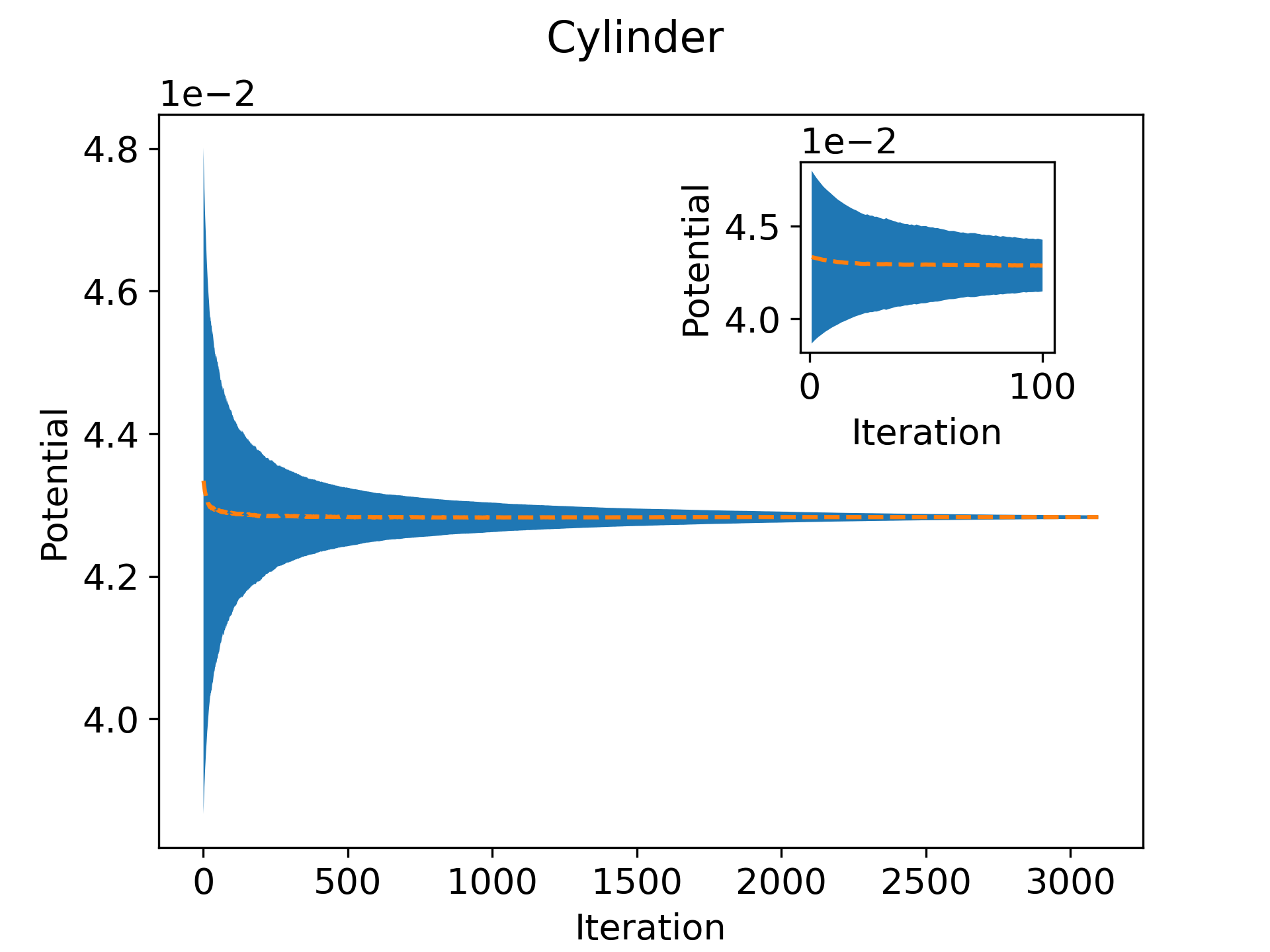}
} &
\subfloat[]{
\includegraphics[width=0.5\textwidth]{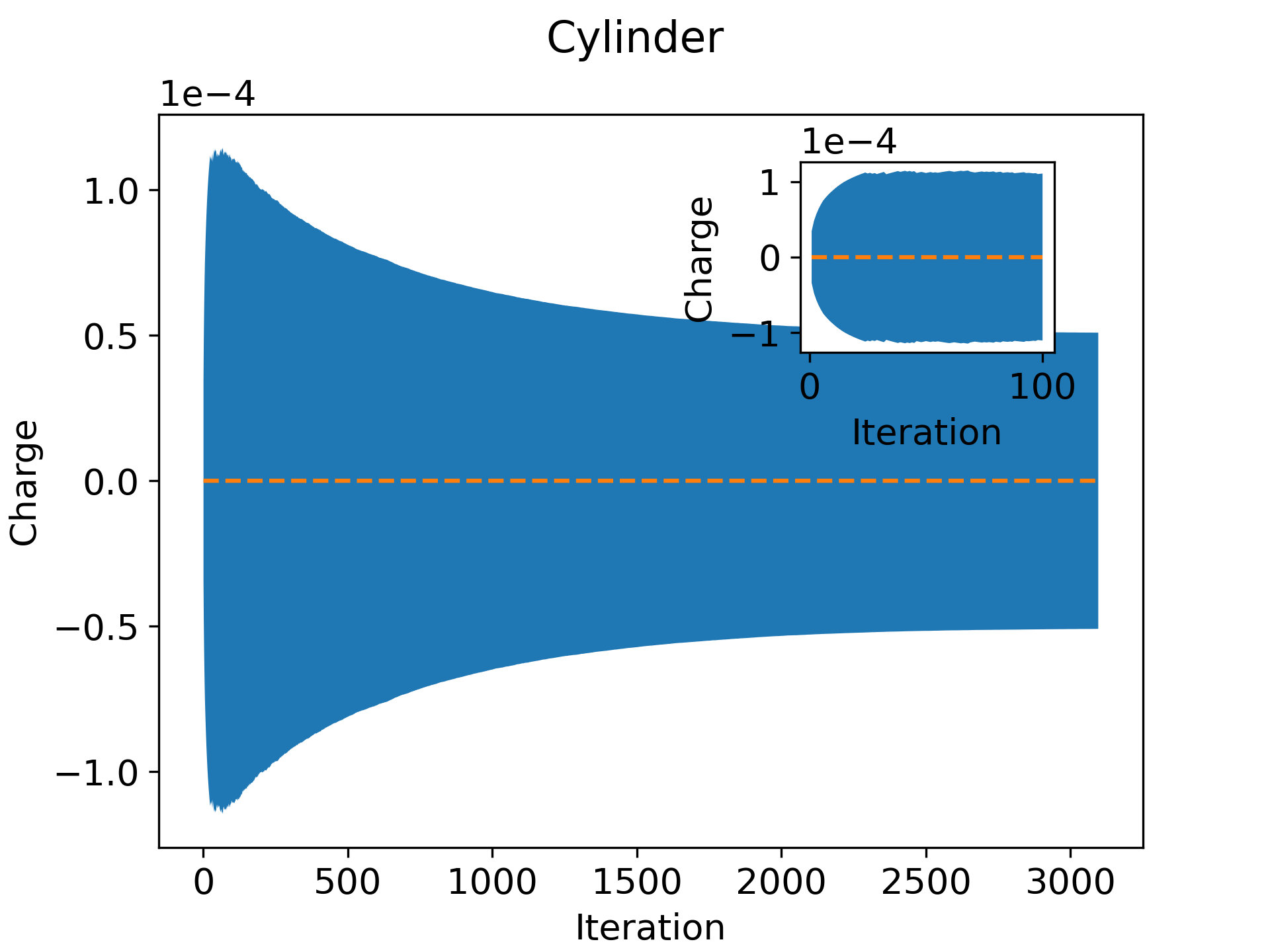}
} \\
\subfloat[]{
\includegraphics[width=0.5\textwidth]{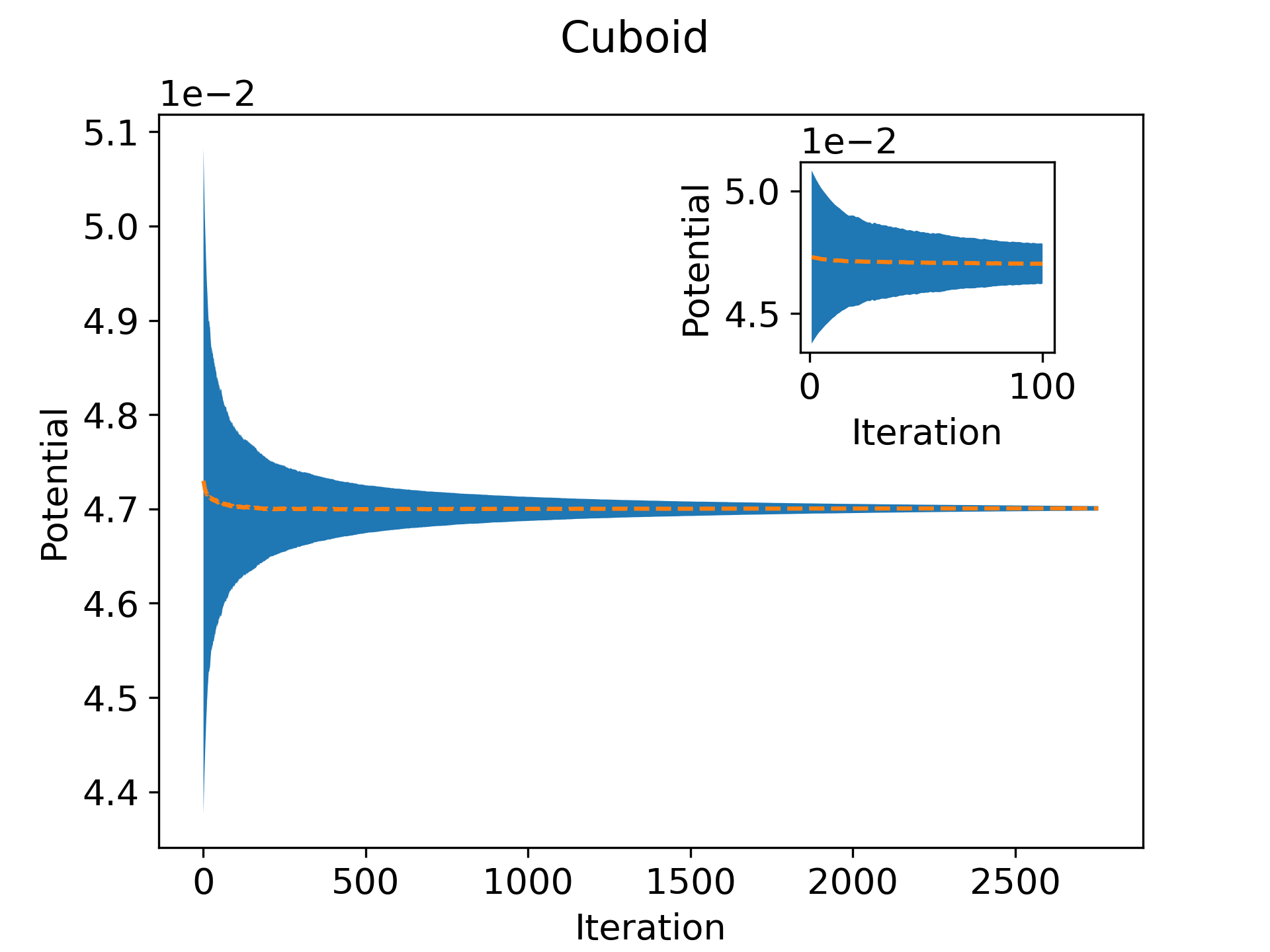}
} &
\subfloat[]{
\includegraphics[width=0.5\textwidth]{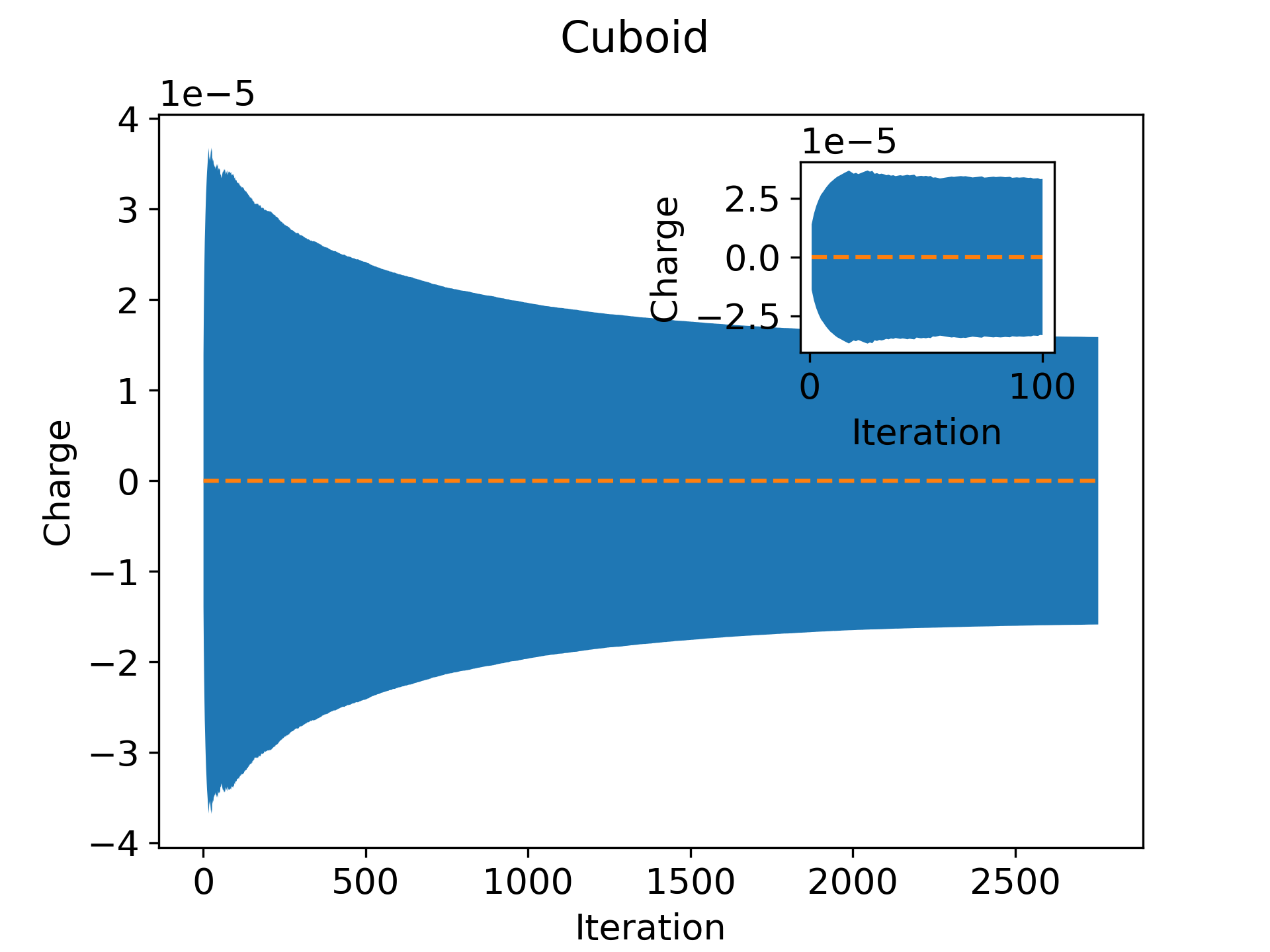}
}
\end{tabular}
\caption{Average and standard deviations over all (main graph) and 100 first iterations (inset graph): a) potential on the cylinder, b) charge on the cylinder, c) potential on the cuboid, d) charge on the cuboid. The geometric parameters of two objects are: cylinder with the base center at $(0, 0, 0)$, base radius $3$ and height $10$; cuboid with the initial vertex at $(0, 0, 0)$ and sides $2, 2, 20$ in the direction of $x, y, z$-axes respectively. A unit point charge is positioned at $(15, 15, 15)$ for both objects.}
\label{Fig:potential_uneven}
\end{figure}

In Fig. \ref{Fig:potential_cubes} we present the electric potential distribution for four distinct configurations of a point charge close to a cube before and after the RH calculation. The initial surface charge distribution at the cube is zero. From the plots in the Figure it is evident that the average initial potential (which fully originates from the external point charge) approximates the final potential very well. Furthermore, in Fig. \ref{Fig:potential_uneven} the convergence of the average electric potential and the surface charge density is presented for a cylinder and a cuboid. The average electric potential changes very little, whereas its standard deviation converges to zero. The surface charge density average remains zero throughout the calculation and its standard deviation converges to a non-zero value. 

\begin{figure}[ht!]
\centering
\begin{tabular}{cc}
\subfloat[]{
\includegraphics[width=0.6\textwidth]{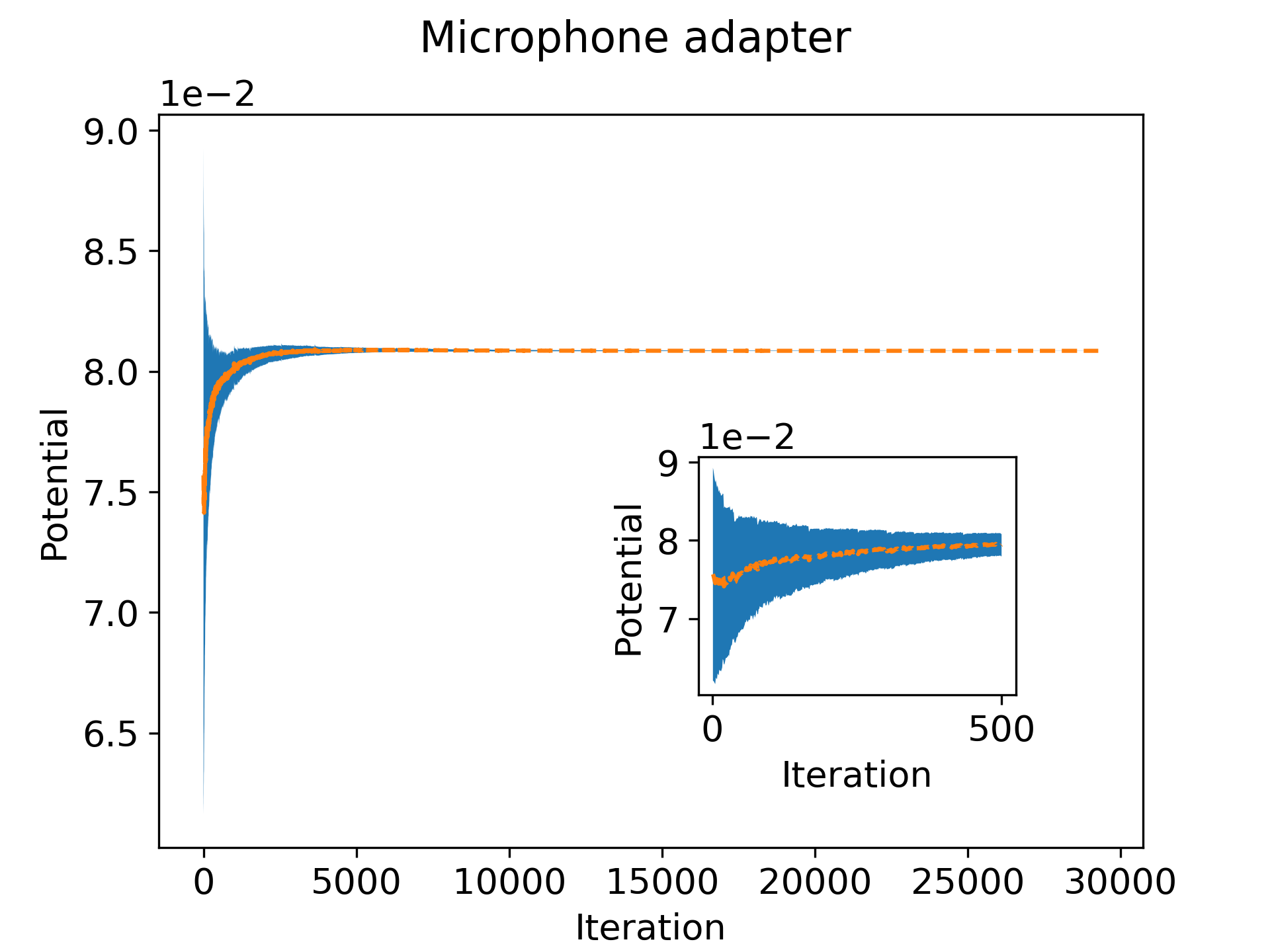}
} &
\subfloat[]{
\includegraphics[width=0.25\textwidth]{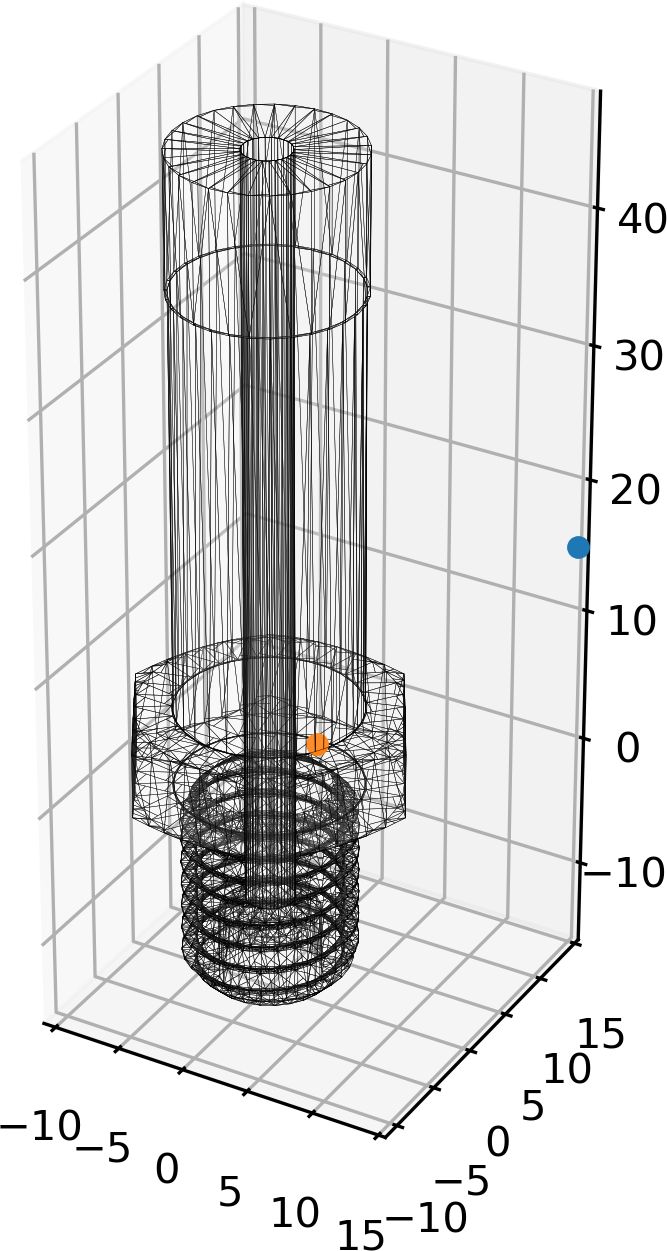}
} \\
\subfloat[]{
\includegraphics[width=0.6\textwidth]{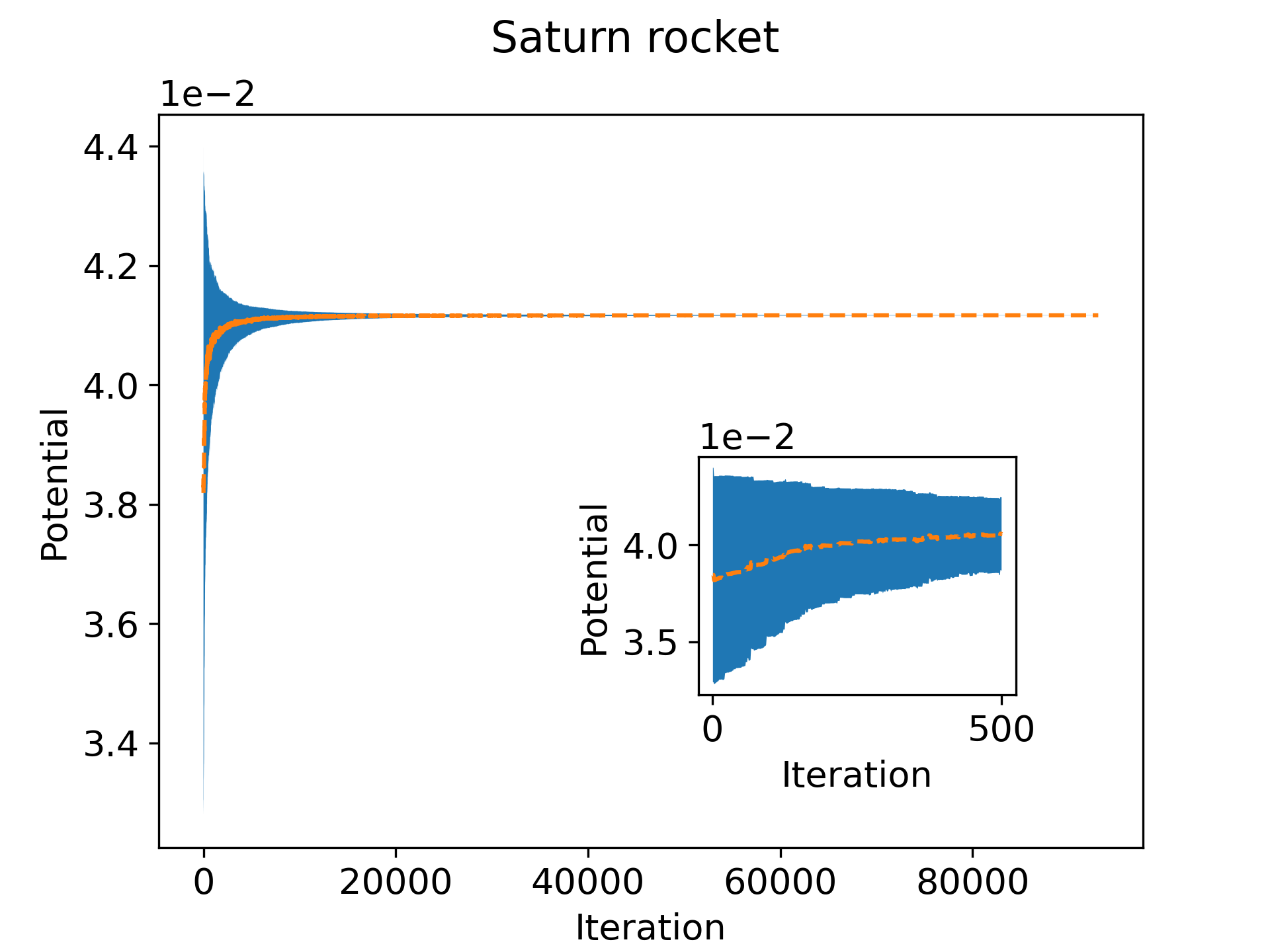}
} &
\subfloat[]{
\includegraphics[width=0.4\textwidth]{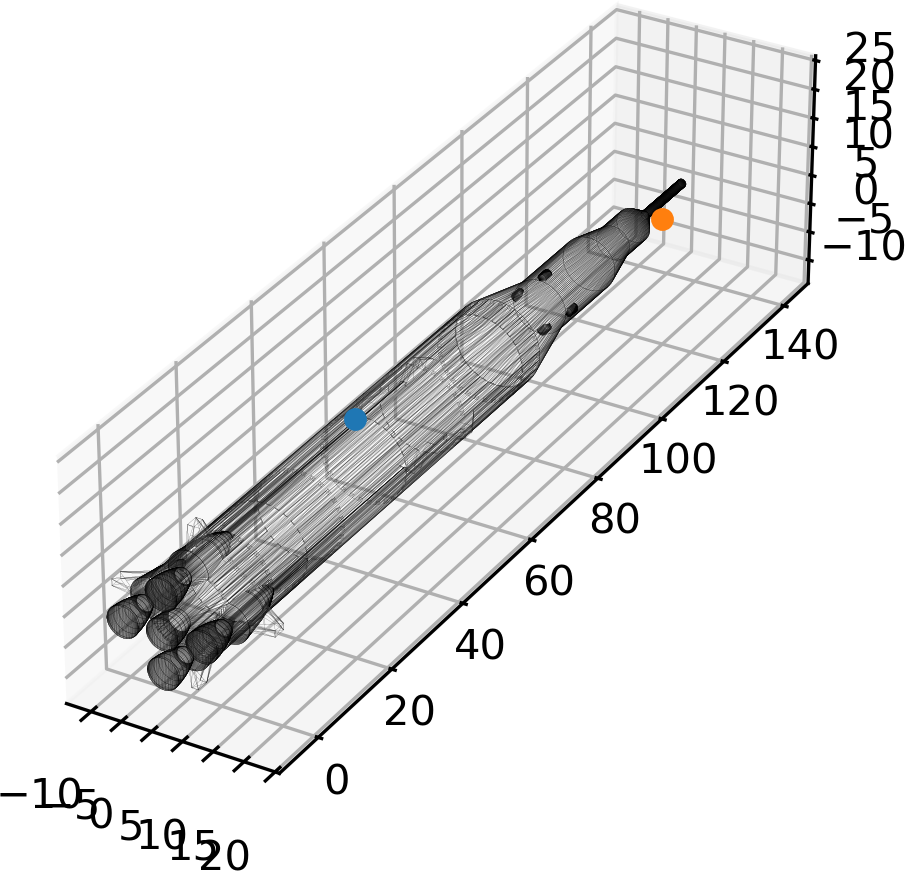}
}
\end{tabular}
\caption{The average and standard deviations of potential over all and 500 first iterations for two objects: (top) microphone adapter  with unit point charges at $(15, 15, 15)$ and $(10, -11, 12)$; (bottom) Saturn rocket model with unit point charges at $(15, 20, 25)$ and $(20, 100, 20)$. Relative difference between initial and final average potential is 5.95\% and 6.55\% for adapter and rocket respectively.}
\label{Fig:uneven_convergence}
\end{figure}

Even for geometrically very uneven objects, differing a lot from the spherical shape, the initial average electric potential is a good approximation of the equilibrium electric potential of an insulated conducting object. In Fig. \ref{Fig:uneven_convergence}, calculations for two realistic objects of very elongated shape with many geometric details are presented: a microphone adapter and a Saturn rocket model. For both objects, the initial average electric potential approximates the equilibrium electric potential with an error below $10\%$. 

The results presented in this subsection show two major tendencies in the approximation of neglecting the third (integral) term in (\ref{eq:potJext}). The more the insulated conducting object is geometrically different from the sphere, the bigger is the difference between the equilibrium electric potential and the average electric potential from the external sources. In addition, this difference is larger if the distance of the external charges from the insulated object is smaller. These tendencies do not express the full complexity of the interplay between the geometry of the insulated conducting object and the positions of external electric charges, but they may serve as a rough approximation of the main determinants of the studied potential difference on the system spatial characteristics (i.e. object geometry and electric charges' positions). 

To measure the spatial characteristics relevant for the two aforementioned tendencies, a number of dimensionless quantities could be used. We first introduce the barycenter of the insulated object surface located at
\begin{equation}
\vec{R}=\frac{1}{S} \int_S \vec{r} \, dS \approx \frac{\sum_i \vec{r}_i \Delta S_i}{\sum_i \Delta S_i} \, .
\label{eq:surfaceCM}    
\end{equation}
Here, $\vec{r}$ refers to the radius vector of points on the surface of the insulated object, $\vec{r}_i$ are radius vectors of the triangle barycenters, and $\Delta S_i$ denote areas of the triangles into which the surface of the insulated object is divided.  
For each point $P$ on the surface (all triangle barycenters), a line through that point and the surface barycenter can be drawn. This line intersects the surface at at least one point on the surface of the object. The distance between the point $P$ and other intersections of the line and the surface of the object is indicated by $d$. The maximum value of $d$ is indicated by $d_{max}$, and the minimal value of $d$ is indicated by $d_{min}$. Furthermore, the distance between the external charge located at $\vec{a}_i$ and the barycenter of the insulated surface is indicated by $d_{q,i}$.

As an indicator of how much the geometry differs from the spherical one, we introduce a dimensionless quantity
\begin{equation}
c_g=\frac{d_{max}}{d_{min}}-1 \, .  
\label{eq:cg}
\end{equation}
For a spherical geometry, $c_g=0$. An indicator of (average) proximity of external charges to the insulated object is introduced as a dimensionless quantity
\begin{equation}
c_c=\frac{1}{N} \frac{2}{d_{max}+d_{min}} \sum_i d_{q,i} \, .
\label{eq:cc}
\end{equation}
The value of  $c_c$ is small for charges close to the insulated object and large for charges far from the insulated object. A systematic analysis of how the accuracy of the approximation depends on parameters such as $c_g$ and $c_c$ is left for future work.  


\begin{figure}[ht!]
\centering
\begin{tabular}{cc}
\subfloat[]{
\includegraphics[width=0.5\textwidth]{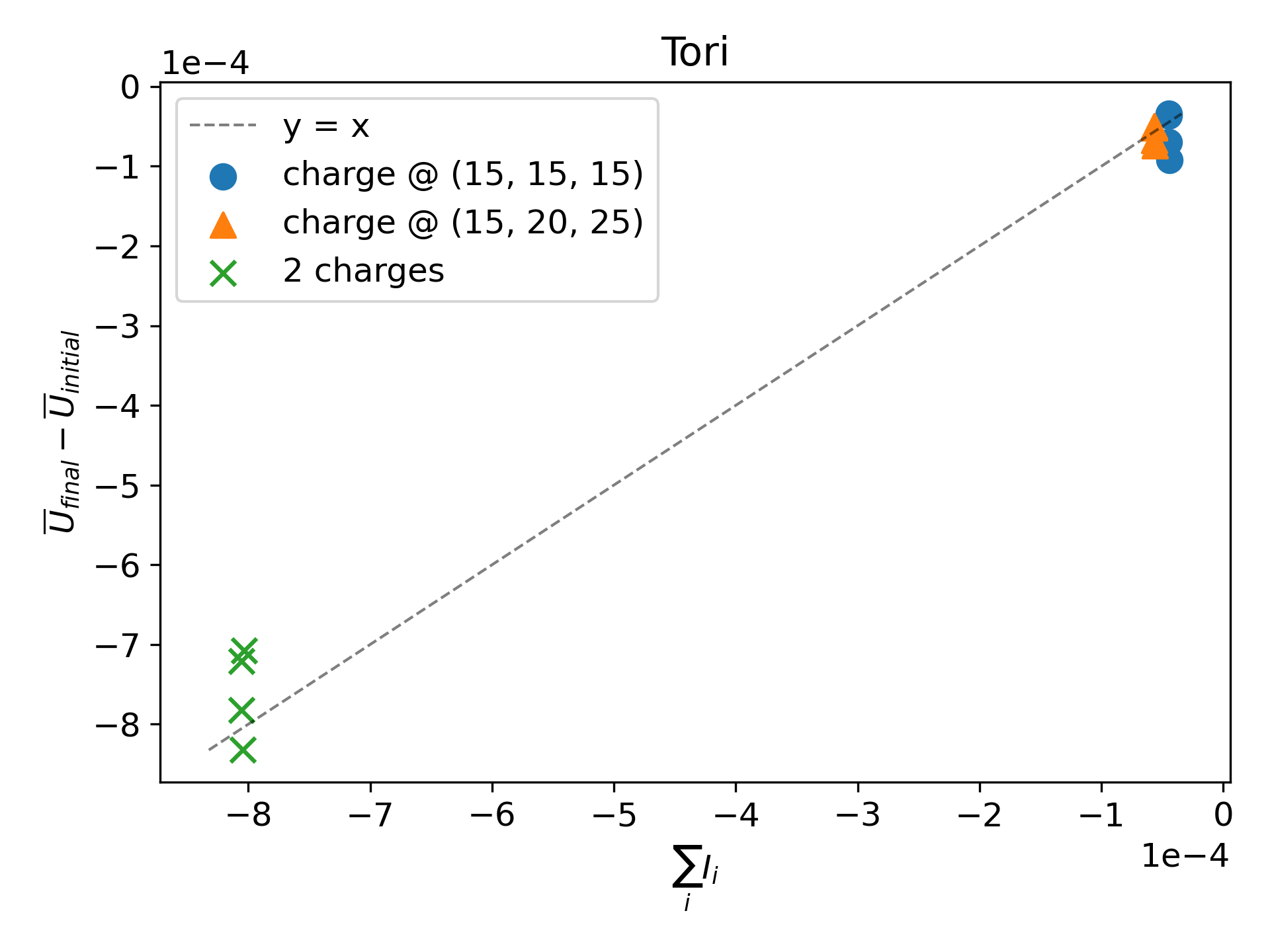}
} &
\subfloat[]{
\includegraphics[width=0.5\textwidth]{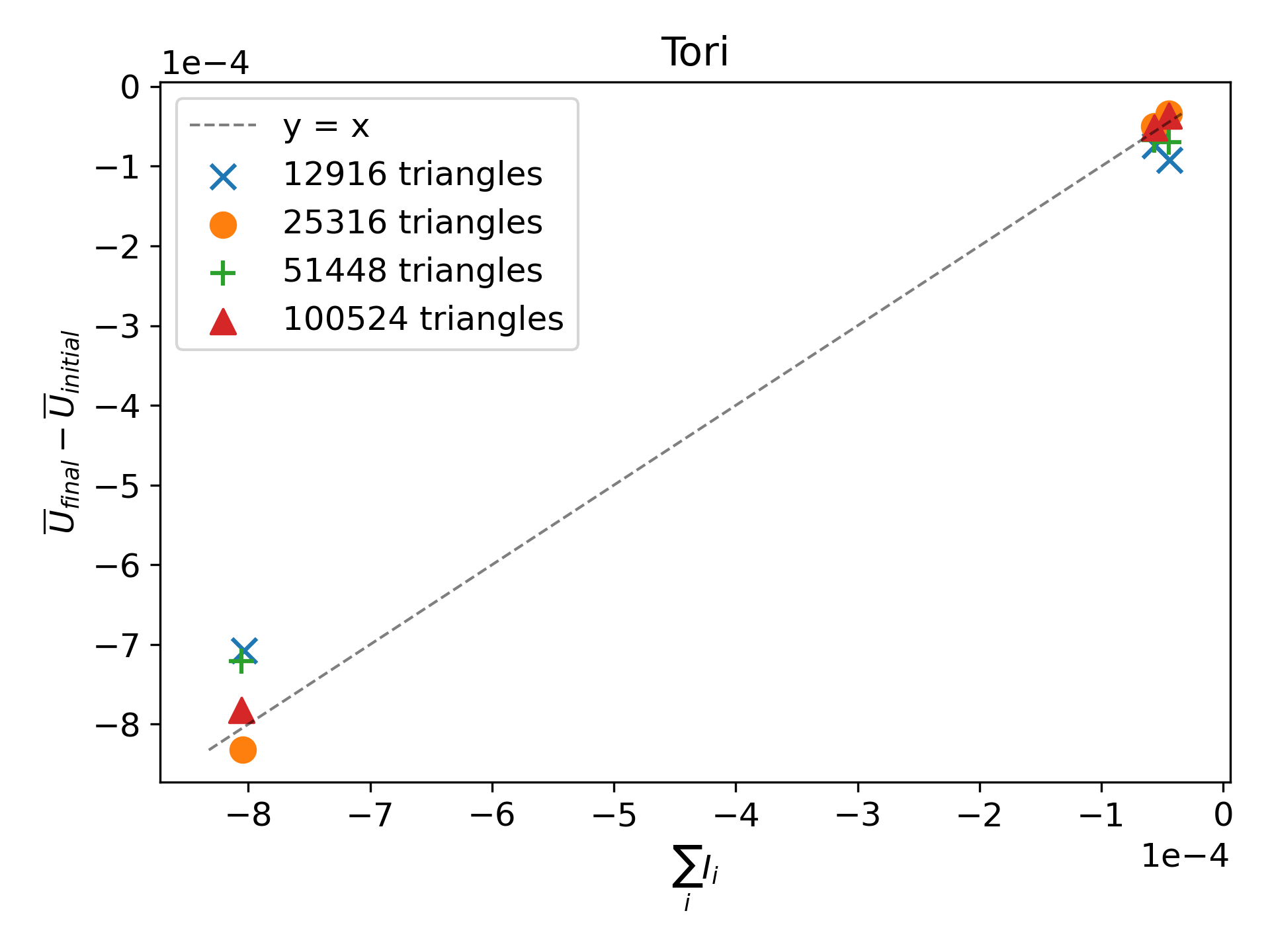}
}
\end{tabular}
\caption{Comparison of calculated integral value and the difference between the final and initial average potentials for torus with radii $R = 10$ and $r = \frac{10}{3}$: a) coloring by different point charges positions, b) coloring by number of torus mesh triangles.}
\label{Fig:integral_check2}
\end{figure}

\begin{figure}[ht!]
\centering
\begin{tabular}{cc}
\subfloat[]{
\includegraphics[width=0.5\textwidth]{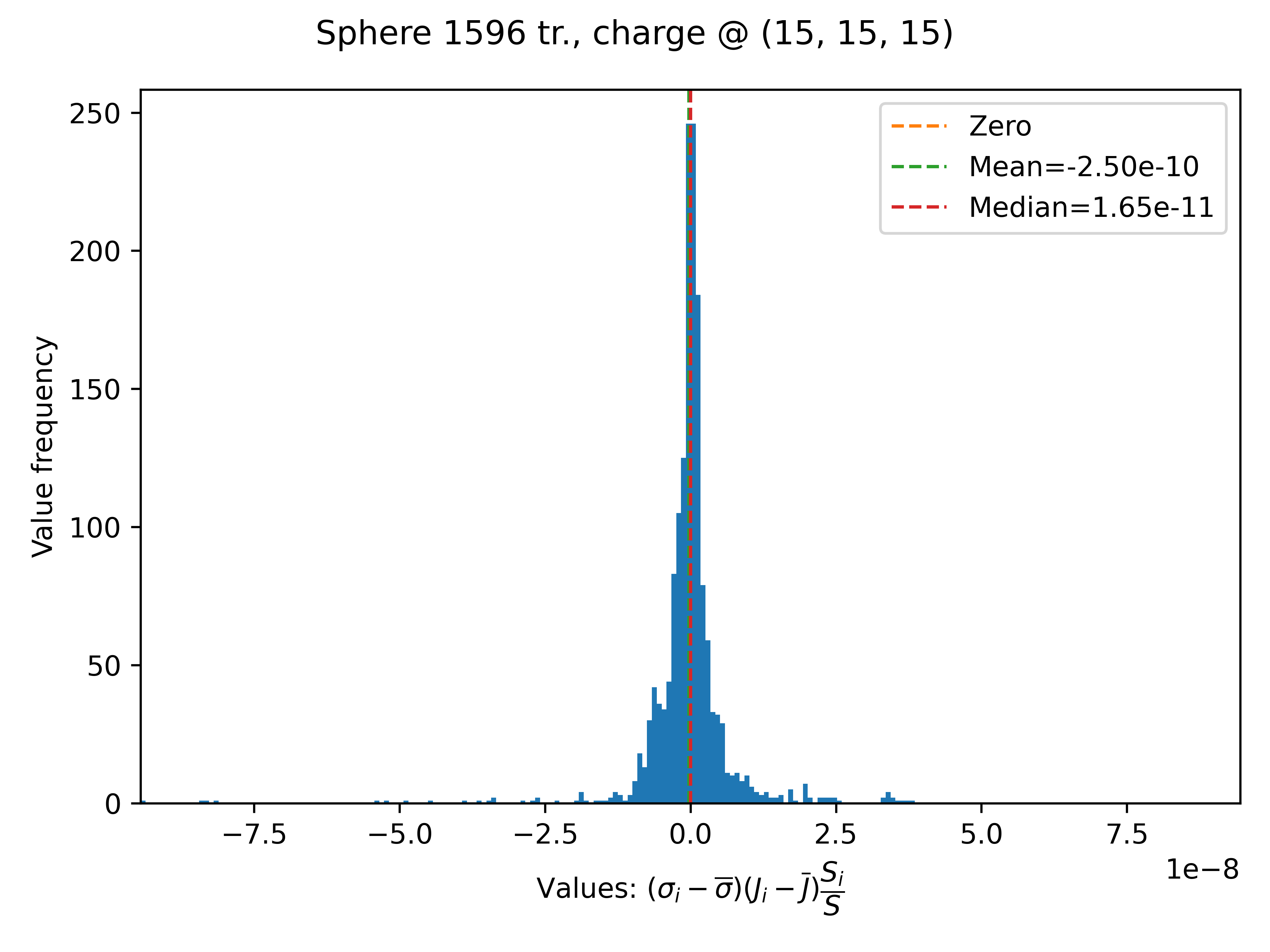}
} &
\subfloat[]{
\includegraphics[width=0.5\textwidth]{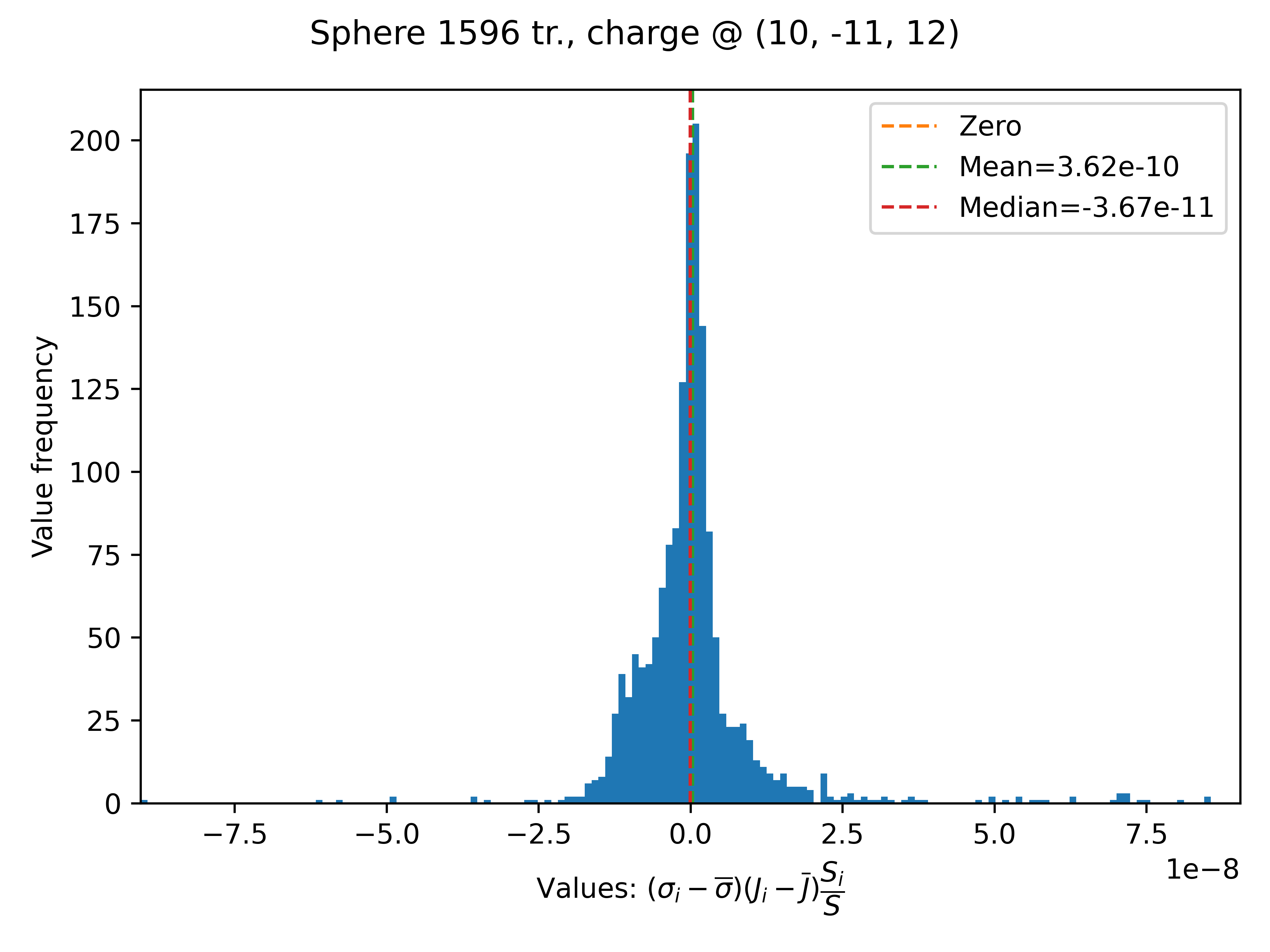}
} \\
\subfloat[]{
\includegraphics[width=0.5\textwidth]{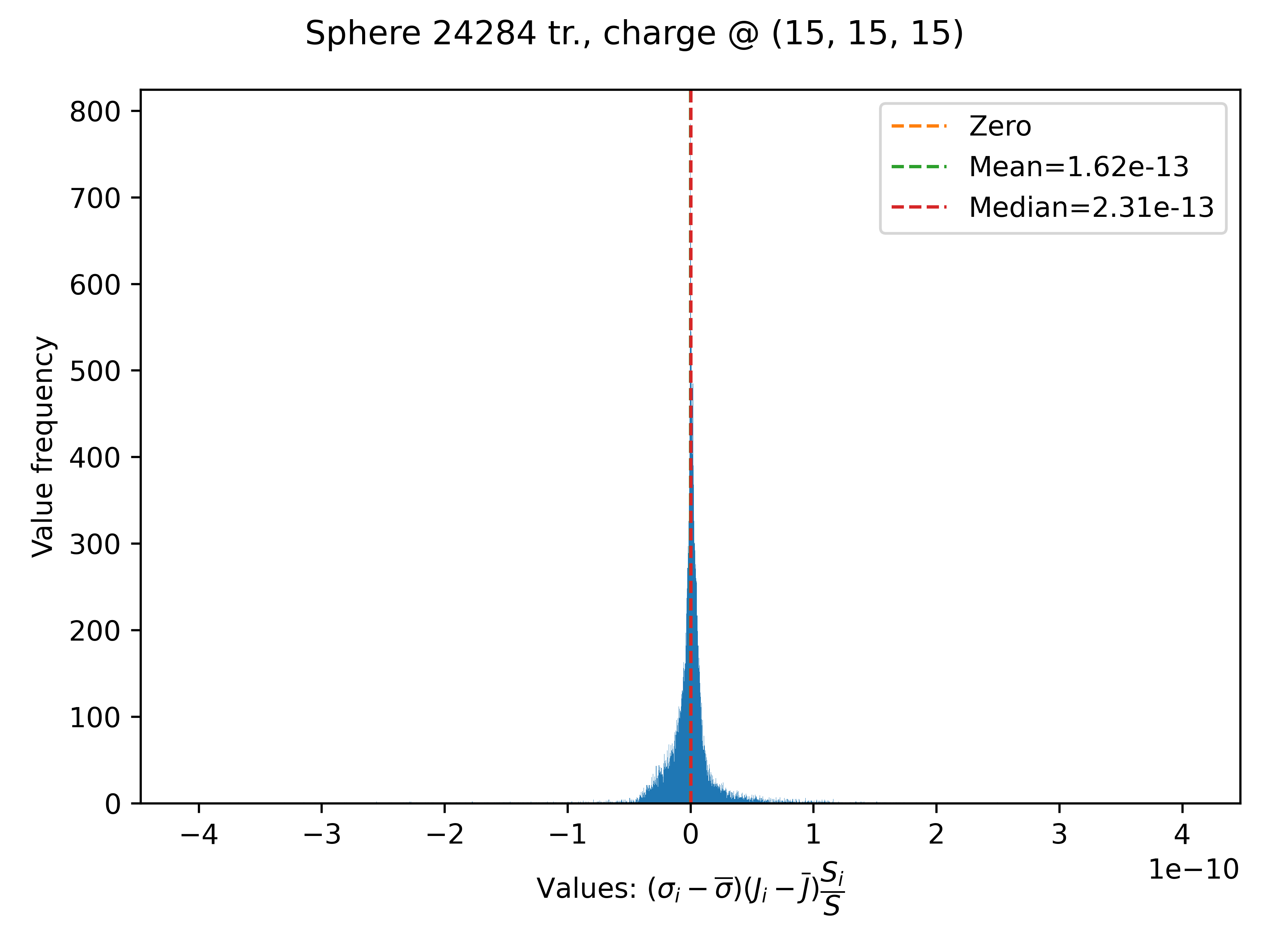}
} &
\subfloat[]{
\includegraphics[width=0.5\textwidth]{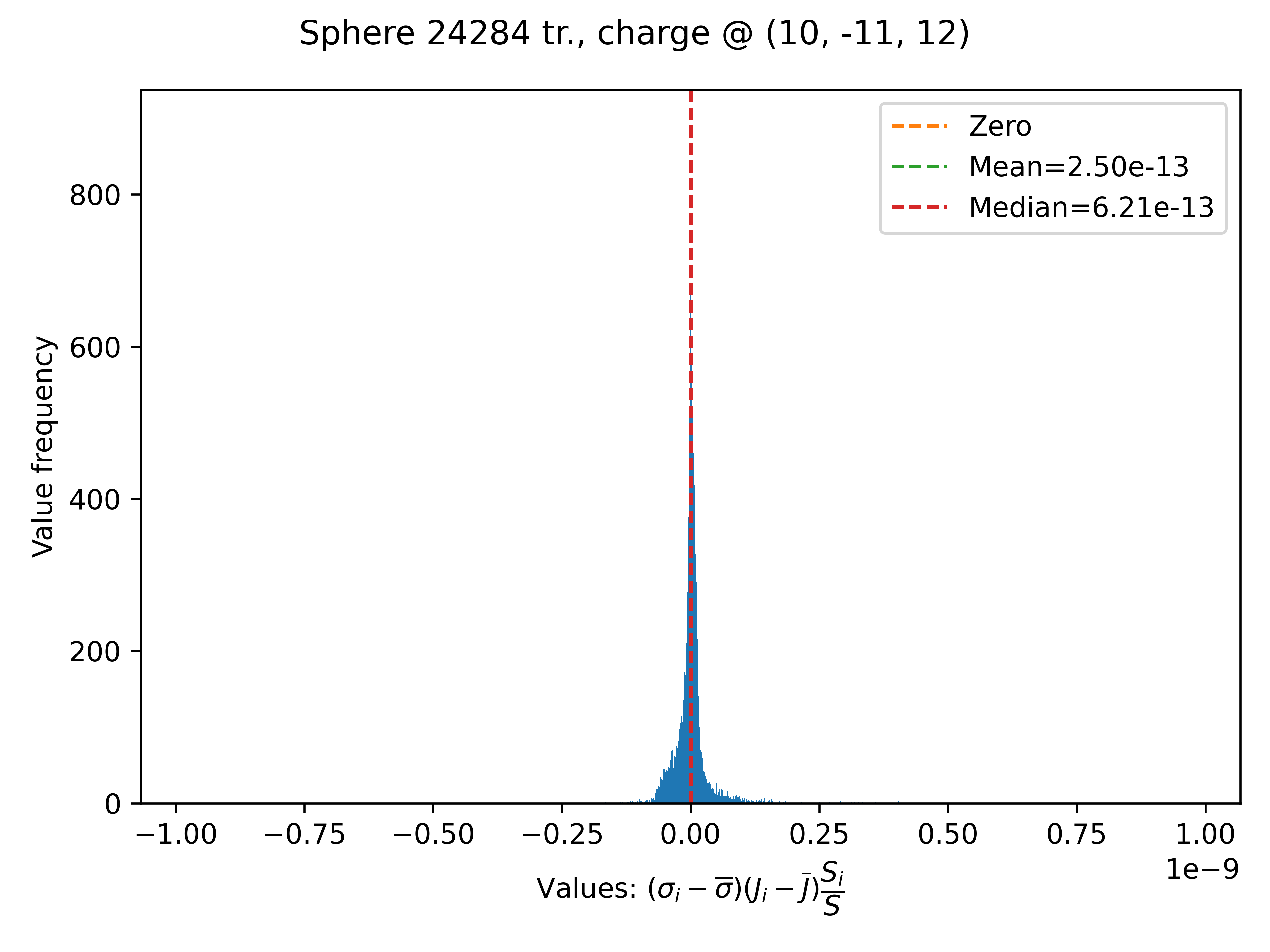}
}
\end{tabular}
\caption{Histograms of integral expression in (\ref{eq:potJext}) for spheres of radius 10 and center at $(0, 0, 0)$ with triangulations: 1596 triangles (top graphs) and 24284 triangles (bottom graphs). Single external unit charge is placed at points: $(15, 15, 15)$ (left graphs) and $(10, -11, 12)$ (right graphs). Dashed lines represent zero, mean and median values (with their numeric value also listed).}
\label{Fig:hist_spheres}
\end{figure}


\begin{figure}[ht!]
\centering
\subfloat[]{
\includegraphics[width=0.5\textwidth]{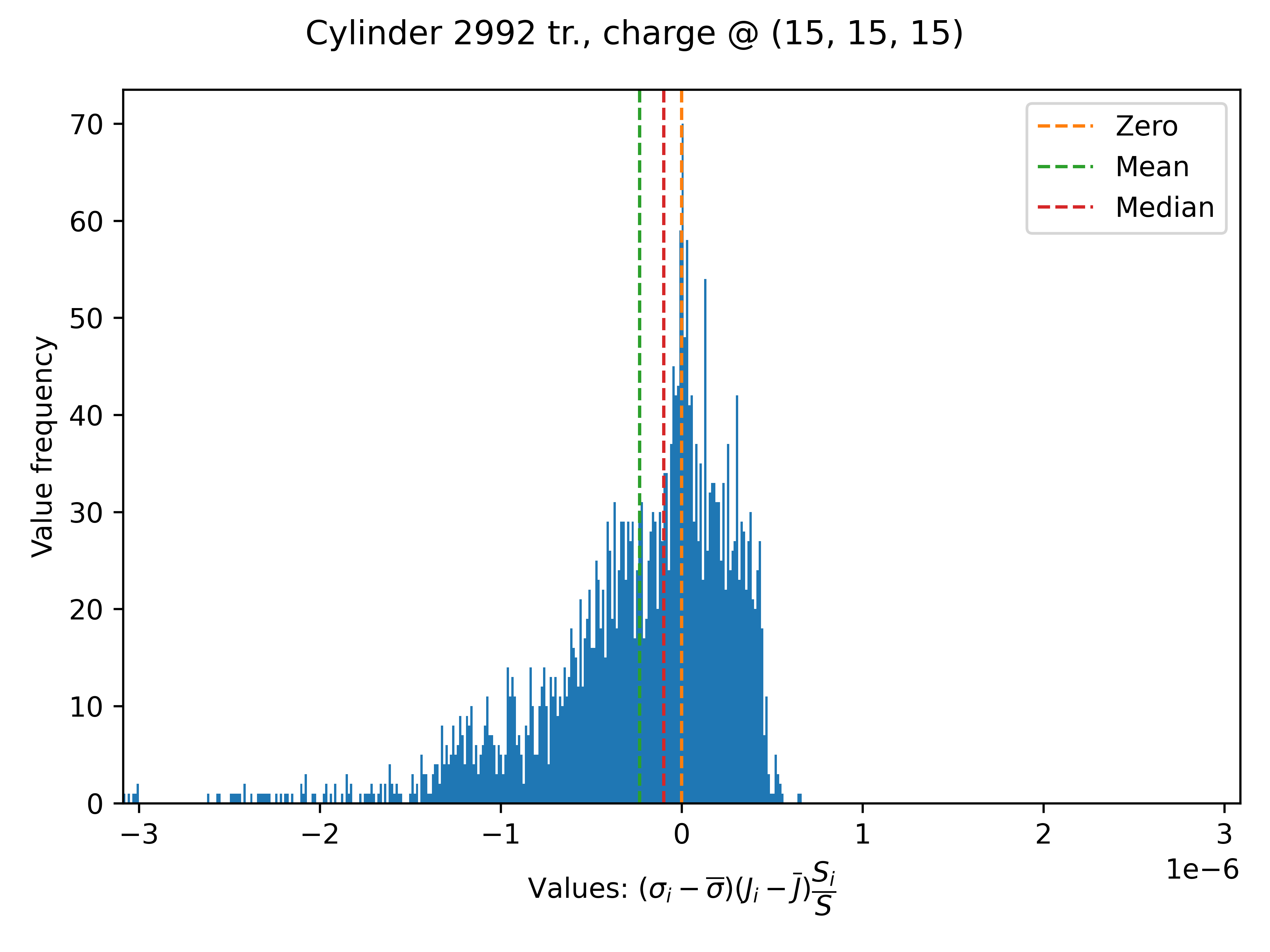}
}
\subfloat[]{
\includegraphics[width=0.5\textwidth]{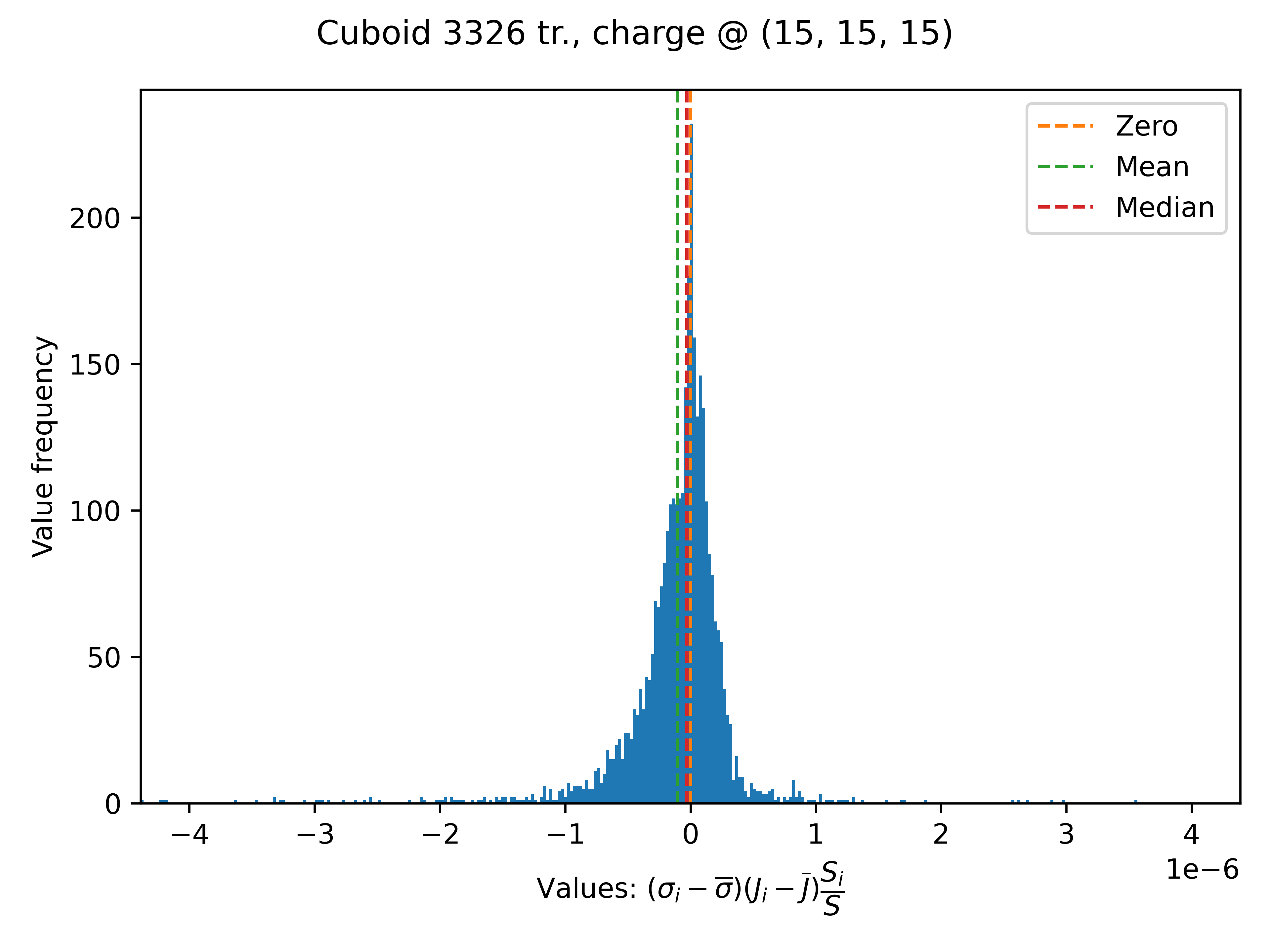}
}
\caption{Histograms of integral expression in (\ref{eq:potJext}) for: a) cylinder with the base center at $(0, 0, 0)$, base radius $3$ and height $10$; b) cuboid with the initial vertex at $(0, 0, 0)$ and sides $(2, 2, 20)$ in the direction of $(x, y, z)$ axes respectively. Both simulations included external unit point charge located at $(15, 15, 15)$. Dashed lines represent zero, mean and median values respectively.}
\label{Fig:hist_solids}
\end{figure}

\begin{figure}[ht!]
\centering
\subfloat[]{
\includegraphics[width=0.5\textwidth]{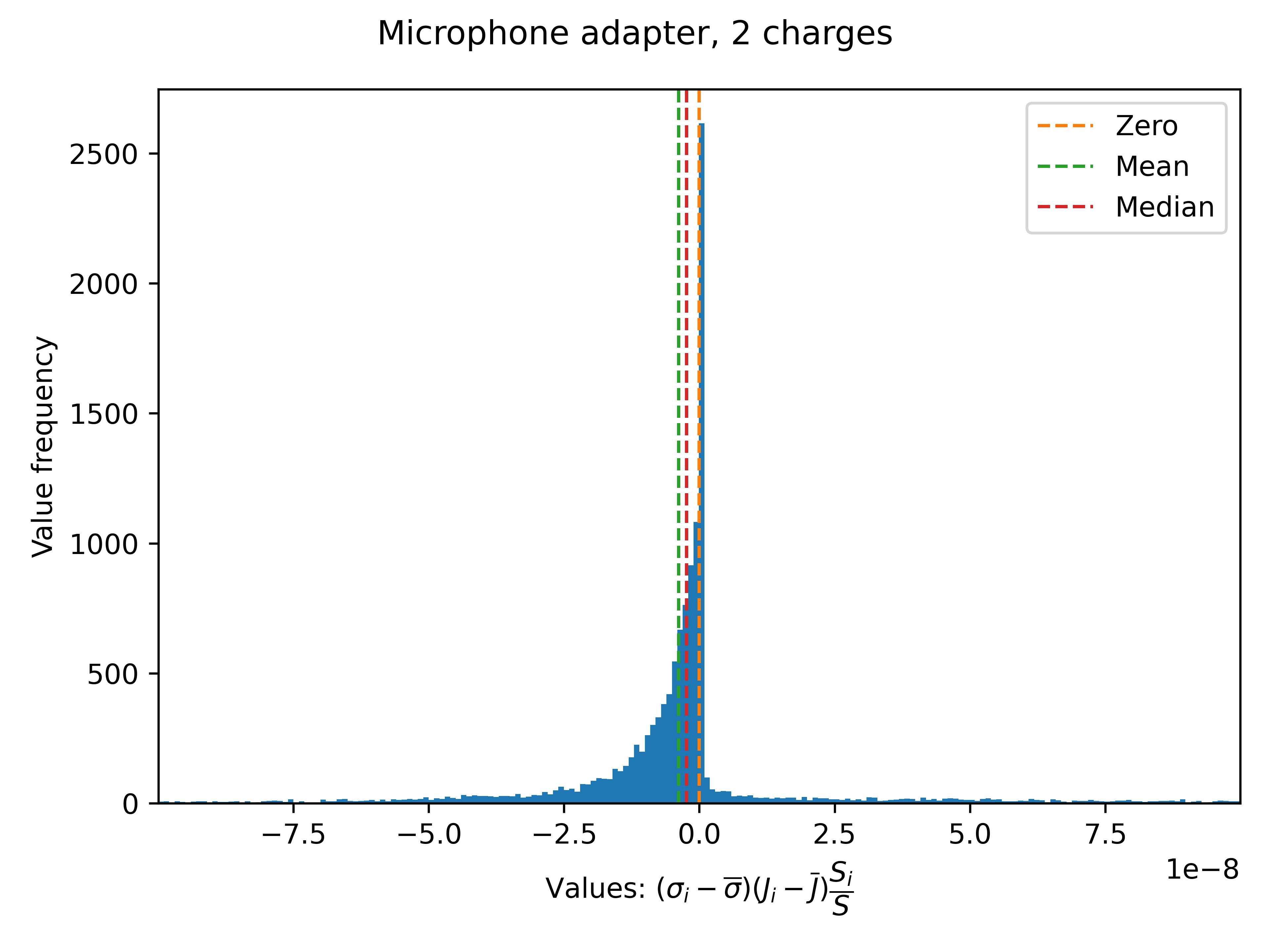}
}
\subfloat[]{
\includegraphics[width=0.5\textwidth]{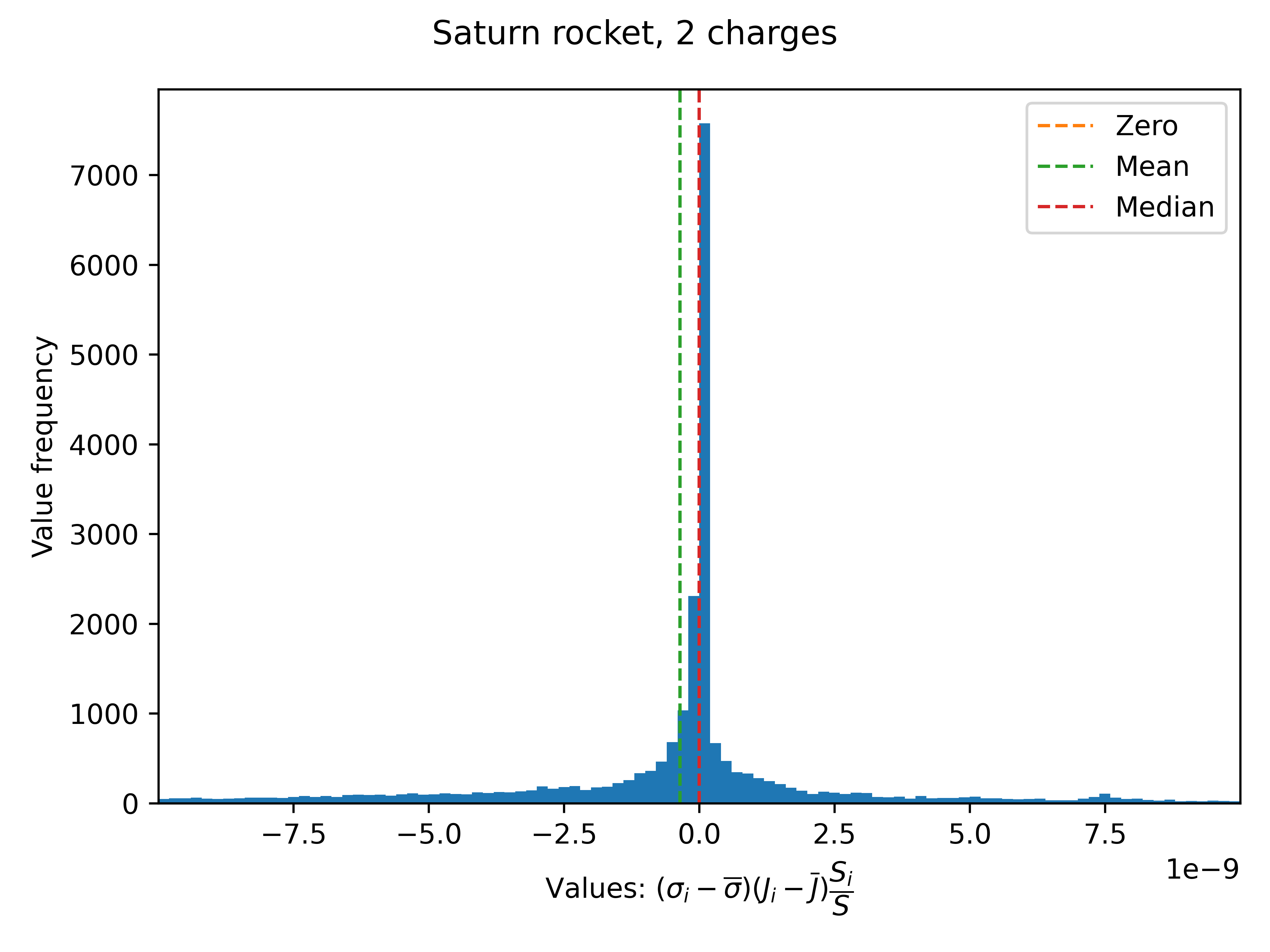}
}
\caption{Histograms of integral expression in (\ref{eq:potJext}) for two objects shown in Fig. \ref{Fig:uneven_convergence}: a) microphone adapter with two external unit point charges placed at $(15, 15, 15)$ and $(10, -11, 12)$; b) Saturn rocket model with unit point charges at $(15, 20, 25)$ and $(20, 100, 20)$. Dashed lines represent zero, mean and median values respectively.}
\label{Fig:hist_objects}
\end{figure}


\subsection{The size of the integral term}

\label{sec:histogram}


The value of electric potential of a charged insulated conducting body is given by the expression (\ref{eq:potJext}). One may perform a consistency test by numerically calculating the electric potential value (at the LHS of (\ref{eq:potJext})) and compare it with the sum of three terms on the RHS of  (\ref{eq:potJext}). Here the value of the third term is also calculated from the results of the RH calculation. It is important to stress that this test is sensitive to issues of precision in the electric potential evaluation, surface discretization and geometry details. To perform such a test, we consider an uncharged torus in presence of external point charges for different torus surface triangulations. In Fig. \ref{Fig:integral_check2} we compare the difference of the final and initial average potential and the third (integral) term in (\ref{eq:potJext}). The obtained results show a very good agreement of the aforementioned quantities, as expected from (\ref{eq:potJext}).

The analytic and numerical results presented so far, demonstrate that this electric potential value can be very well approximated by the first two terms in (\ref{eq:potJext}), i.e. that the third term in this expression is (in most cases) small. As this term is an integral, its smallness may have different causes. It is possible that the integrand is in some regions of integration (surface of the insulated object) is negative and in others positive and that these contributions largely cancel (cancellation effect). Another possibility is that the integrand is predominantly of the same sign over the integration region, but its size is small (small size effect). To investigate this problem in more detail, we produce a histogram of the function $\frac{S_i}{S}(\sigma(\vec{x}_i)-\langle \sigma \rangle) (J(\vec{x}_i) - \langle J \rangle)$ for the triangles of several objects presented in Figures \ref{Fig:hist_spheres}, \ref{Fig:hist_solids}, and \ref{Fig:hist_objects}. All histograms in these Figures show distributions with mean and median close to zero. For spheres in Fig. \ref{Fig:hist_spheres} the distributions are quite symmetric and get narrower for the more refined meshes. The near symmetry of the distributions points to complete cancellation of negative and positive contributions in the integral, as expected from theoretical considerations for spheres. For non-spherical objects, such as those Figs \ref{Fig:hist_solids} and \ref{Fig:hist_objects}, the distributions are skewed and the cancellation exists, but it is only partial. A broad general conclusion is that the third term in (\ref{eq:potJext}) is generally small owing to the cancellation of various contributions to the integral and this cancellation is smaller (and the integral term more significant) for objects that deviate more from the spherical shape.   


\section{Discussion}


The results presented in the preceding sections show that the electric potential of insulated objects can be calculated exactly or efficiently approximated in the $J$ formalism  introduced in this paper. The approximation consists in keeping only the first two terms in (\ref{eq:potJext}) and disregarding the third, integral term. Analytical investigations of the neglected integral term indicate that it is small (i.e. the approximation is good) when the distribution of $J$ values is close to its surface average $\langle J \rangle$, although the full answer lies in the interplay of $J$ values and charge surface density, as visible from the integral term in (\ref{eq:potJext}). Generally, the more objects deviate from sphericity, the weaker is the aforementioned approximation.    

As already stated, the practical advantage lies in the much simpler calculation required for the approximation of the insulated object potential. For external charges, it is only necessary to calculate the average potential of these charges on the surface of the object. This calculation depends on both the object geometry and the positions of the charges. For the case of a charged insulated object, in addition to the charge $Q$, it is only necessary to know the average value $\langle J \rangle$ and the surface of the object $S$. For many standard conductor geometries, these values can be calculated (analytically or numerically) and tabulated. This further simplifies and facilitates the applications. Concrete applications might be expected from the macroscale (e.g. in energy transmission or electric machinery) to the microscale (MEMS or nanoparticles). 

The limitations of the approach can be assessed with confidence only after the study of many insulated object geometries and charge configurations. Currently it seems that the approximation of the potential starts to be less precise when the external charges are close to the object and when the object is elongated (i.e. less spherical). It is however, both surprising and encouraging that even for very elongated objects the approximation works within tens of percent.   

Real systems seldom contain just a single insulated conducting object. It is therefore natural to ask if the approach presented in this paper can be extended to systems with multiple insulated conducting objects, objects at fixed external potentials, or dielectrics. Here, the situation is considerably more complicated since the charge distribution on one object is influenced by the very presence of other objects. The extension of this approach will depend both on the deeper understanding of why the approach works for single insulated conducting objects and on dedicated numerical analyses of systems with multiple conducting or dielectric objects.  

Given that the electric potential can be expressed as a sum of several terms, of which some are dominant and some negligible or considerably suppressed, it is also intriguing to ask if the $J$ formalism could be related to some expansion based approaches, such as multipole expansion \cite{Jackson}. Although this question is at the moment completely open, should such an expansion approach exist, the expansion parametetrs worthy of investigation might be $c_g$ and $1/c_c$ introduced in (\ref{eq:cg}) and (\ref{eq:cc}).  






A reasonable question is whether the approximation scheme presented in this paper could be successfully applied to an even more important problem of conductor(s) kept at fixed electric potentials. Though this question merits a dedicated and detailed numerical study, let us illustrate the usefulness of the said approach on an example of a single conducting object kept at a fixed potential. In this example we assume that there are no external charges and the only relevant charge distribution is present on the surface of the conducting body. The electric potential at the point on the surface $S$ of the conducting body located at $\vec{x}$ is:   
\begin{equation}
\phi (\vec{x})= \int_{S} k \frac{\sigma (\vec{x'})}{|\vec{x}-\vec{x'}|} d S' \, . 
    \label{eq:fixed_potential}
\end{equation}
By the definition of the problem, $\phi (\vec{x})= U$, where $U$ is the value of the electric potential at which the conducting object is kept. Following the same averaging procedure as in the case of an insulated body, the average electric potential can be written as
\begin{equation}
\langle \phi \rangle =  \frac{1}{S} \int_S d S' \sigma(\vec{x'}) J(\vec{x'}) \, 
\label{eq:fixed_potJ}    
\end{equation}
and more conveniently expressed as
\begin{equation}
\langle \phi \rangle = \langle \sigma \rangle \langle J \rangle + \frac{1}{S} \int_S d S' (\sigma(\vec{x'})-\langle \sigma \rangle) (J(\vec{x'}) - \langle J \rangle) \, .
\label{eq:fixed_potJext}    
\end{equation}
Taking into consideration that the potential at the surface of the conducting object is fixed, one obtains
\begin{equation}
U = \frac{Q}{S} \langle J \rangle + \frac{1}{S} \int_S d S' (\sigma(\vec{x'})-\langle \sigma \rangle) (J(\vec{x'}) - \langle J \rangle) \, ,
\label{eq:fixedU}    
\end{equation}
where $Q$ is the total charge on the surface of the conducting object. This expression leads to a potentially useful relation for the electric capacity $C$ of a body maintained at a fixed electric potential:
\begin{equation}
\frac{1}{C} \equiv \frac{U}{Q} = \frac{\langle J \rangle}{S} + \frac{1}{S} \int_S d S' \frac{\sigma(\vec{x'})-\langle \sigma \rangle}{Q} (J(\vec{x'}) - \langle J \rangle) \, .
\label{eq:def_capacity}    
\end{equation}
If the integral term in (\ref{eq:def_capacity}) is small compared to the leading term $\frac{\langle J \rangle}{S}$, then a useful approximation for the electric capacity follows:
\begin{equation}
C \approx \frac{S}{\langle J \rangle} \, .
\label{eq:approx_capacity}    
\end{equation}
For a sphere, the integral in (\ref{eq:def_capacity}) vanishes owing to the constancy of $J$ on the surface of the sphere and the expression (\ref{eq:approx_capacity}) produces a well-known result for the capacity $C=4 \pi \varepsilon_0 R$ (with $k=\frac{1}{4 \pi \varepsilon_0}$).
The main advantage of this approximation is that it relies only on the geometry and requires no calculations for the equilibrium surface charge distribution. 
Although preliminary investigations of the usefulness of formula (\ref{eq:approx_capacity}) for other geometries are encouraging, its systematic study is required and it is left for future work \cite{Cap}. The expression (\ref{eq:approx_capacity}) has some similarity to the result of \cite{Chow} that the capacitance is strongly dependent on the surface of the conductor (more precisely the square root of its surface) and much less on the shape of the conductor. It is an interesting challenge for future work to determine if the $J$ formalism can provide some additional insight into this long-standing result of \cite{Chow}.

\section{Summary and conclusions}

The electric potential of an insulated conducting object in the presence of some arbitrary fixed charge distribution is an important quantity both theoretically and practically. For a general geometry of the insulted object, the determination of this potential requires a numerical finding of the charge distribution at the surface of the object (or some other numerical procedure of comparable complexity). In this paper, we demonstrate that a computationally much simpler alternative exists, and that for a range of insulated object geometries and external charge configurations, it gives a very good approximation of the electric potential value. This alternative approach consists in calculation of the average potential that the external charges produce at the surface of the insulated conducting object. This approach is of purely geometrical nature and we introduce a novel formalism of its description based on a function $J(\vec{x})$ which corresponds to an electric potential that the insulated object with a unit surface charge distribution would produce at $\vec{x}$ on $S$.   

The average electric potential of external charges on the surface of an insulated conducting sphere is actually equal to the electric potential of the insulated conducting sphere in the presence of external electric charges. This result is analytically proven in a novel way using a function $J(\vec{x})$, and its validity is supported by numerical calculations using the Robin Hood method. Although analytical results for the sphere have been obtained (or can be easily obtained) using method of images, these results have been obtained in a new way.

The main practical contribution of this paper lies in the application of this approach to non-spherical objects. The average value of the electric potential of external charges proves to be an excellent approximation of the equilibrium electric potential for various non-spherical geometries and external charge configurations. Although a more systematic study is needed, the quality of the approximation decreases as the object deviates from the spherical shape and as external charges are placed closer to the insulated object surface. However, even in several cases of complicated geometries very different from a sphere (e.g. very elongated), the relative potential difference $\frac{\phi - \langle \phi_{ext} \rangle}{\phi}$ is at most several tens percent. These initial results suggest that the introduced approach could safely give order of magnitude estimates of the electric potential of insulated conducting objects without surface charge calculation for all practically important situations. Verification of this possibility requires extensive future simulations of various geometries and external charge configurations.

From a formal point of view, the reason why the $J$ formalism is so successful lies in the smallness of the third (integral) term in (\ref{eq:potJext}). The histograms presented in section \ref{sec:histogram} provide an insight into a cancellation mechanism making this term small. Indeed, the distributions are centered close to zero with a substantial cancellation of positive and negative contributions. 

An intriguing question is how successfully the approach adopted in this paper could be extended to problems beyond insulated conducting objects. One promising application beyond insulated objects is the calculation of electric capacities of conducting objects of arbitrary geometry. In the Discussion section it is shown how the electric capacity of the conducting object can be approximated using the average value of the $J$ function. A systematic investigation of this promising research avenue is left to future work \cite{Cap}. 

In conclusion, quantities of importance in electrostatic problems, such as electric potential of an insulated conducting object or the electric capacity of a conducting object of arbitrary geometry can be well approximated without determination of the equilibrium surface charge distribution. A geometrical quantity $J(\vec{x})$, interpretable as a potential that a unit charge distribution would produce at a point $\vec{x}$ on the surface of the conducting object, plays a central role in the approach behind this approximation. The results presented in this paper are practically relevant because they demonstrate an efficient approximation scheme for some important quantities in electrostatics. A theoretical contribution is in understanding why this approach works which opens way for possible new applications in electrostatics and beyond.


\begin{thebibliography}{}

\bibitem{Jackson} J. D. Jackson, Classical Electrodynamics (Third Edition), John Wiley and Sons, Hoboken (1999).

\bibitem{Landau} L. D. Landau, E. M. Lifshitz, Electrodynamics of Continuous Media (Vol. 8), Pergamon Press, Oxford (2020).

\bibitem{Griffiths} D. J. Griffiths, Introduction to Electrodynamics, Cambridge University Press (2023).

\bibitem{Reitz} J. R. Reitz, F. J. Milford, R. W. Christy, Foundations of Electromagnetic Theory (Fourth Edition), Pearson Addison-Wesley, San Francisco (2009).

\bibitem{Smythe} W. R. Smythe, Static and Dynamic Electricity (Second Edition), McGraw-Hill Book Company, New York (1950).

\bibitem{HVI} P. Llovera-Segovia, Journal of Electrostatics {\bf 137} (2025) 104118.

\bibitem{propulsion} D. R. Jovel, M. L. R. Walker, and D. Herman, Journal of Propulsion and Power, {\bf 38} (2022) 1051-1081.

\bibitem{micronano} T, Cai, Y. Fang, Y. Fang, R. Li, Y. Yu, and M. Huang, Beilstein J. Nanotechnol. {bf 13} (2022) 390-403.

\bibitem{materials} D. Stamopoulos, Materials {\bf 17} (2024) 5046.

\bibitem{chem1} E. Besley, Acc. Chem. Res. {\bf 56} (2023) 2267-2277. 

\bibitem{chem2} Zh. Long, J. Meng, L. R. Weddle, P. E. Videla, J. P. Menzel,
D. G. A. Cabral, J Liu, T. Qiu, J. M. Palasz, D. Bhattacharyya, C. P. Kubiak, V. S. Batista, and T. Lian, 
Chem. Rev. {\bf 125} (2025) 1604-1628.

\bibitem{agri} J. Cao, Zh. Jin, J. He, G.Ju, L. Mi, Y. Gao, R. Lei, and G. Cheng, Micromachines {\bf 16} (2025) 1285.

\bibitem{tribo} L. Huang, G. Huang, D. Zhang, X. Chen, Friction {\bf 13} (2) (2025) 9440893.

\bibitem{float} P. Llovera-Segovia, P. Molini\'{e}, V. Fuster-Roig, A. Quijano-L\'{o}pez, Journal of Electrostatics {\bf 132} (2024) 103986.

\bibitem{RH1} P. Lazi\'{c}, H. \v{S}tefan\v{c}i\'{c}, H. Abraham, J. Comp. Phys. {\bf 213} (2006) 117.

\bibitem{RH2} P. Lazi\'{c}, H. \v{S}tefan\v{c}i\'{c}, H. Abraham, Eng. Anal. Boundary Elem. {\bf 32} (2008) 76.

\bibitem{RH3} P. Lazi\'{c}, D. Dujmi\'{c}, J. A. Formaggio, H. Abraham, H. \v{S}tefan\v{c}i\'{c}, JINST {\bf 6} (2011) P12003.

\bibitem{RH4} J. A. Formaggio, P. Lazi\'{c}, T. J. Corona, H. \v{S}tefan\v{c}i\'{c}, H. Abraham, F. Gluck, Progress In Electromagnetics Research B {\bf 39} (2012) 1.

\bibitem{Cap} K. Filipan, H Štefančić, in preparation.

\bibitem{Chow} Y. L. Chow, M. M. Yovanovich, J. Appl. Phys. {\bf 53} (1982) 8470.

\end{thebibliography}
\end{document}